\documentclass[12pt]{article}

\usepackage{graphics,graphicx}
\usepackage{amsmath}
\usepackage{mathtools}
\usepackage{epsf}
\usepackage{amsfonts}
\usepackage{amssymb}
\usepackage[english]{babel}
\usepackage[applemac]{inputenc}            % Mac
\usepackage{color}
\usepackage{multirow}
\usepackage{bm}
\usepackage{enumitem}
\usepackage[cal=boondoxo]{mathalfa}
\usepackage{caption}

\date{}

\begin{document}

\title{\bf Wave polarisation and dynamic degeneracy in a chiral elastic lattice}

\author{ G. Carta$^{1}$, I.S. Jones$^{1}$, N.V. Movchan$^{2}$, A.B. Movchan$^{2}$ \\
\small{ $^1$ Liverpool John Moores University, Mechanical Engineering} \\
\small{and Materials Research Centre, Liverpool, L3 3AF, UK} \\
\small{$^2$ University of Liverpool, Department of Mathematical Sciences,} \\
\small{Liverpool, L69 7ZL, UK} \\
\small{Email address: giorgio\_carta@unica.it} }

\maketitle

\begin{abstract}
This paper addresses fundamental questions arising in the theory of Bloch-Floquet waves in chiral elastic lattice systems. This area has received a significant attention in the context of ``topologically protected'' waveforms. Although practical applications of chiral elastic lattices are widely appreciated, especially in problems of controlling low-frequency vibrations, wave polarisation and filtering, the fundamental questions of the relationship of these lattices to classical waveforms associated with longitudinal and shear waves retain a substantial scope for further development. The notion of chirality is introduced into the systematic analysis of dispersive elastic waves in a doubly-periodic lattice. Important quantitative characteristics of the dynamic response of the lattice, such as lattice flux and lattice circulation, are used in the analysis along with the novel concept of ``vortex waveforms'' that characterise the dynamic response of the chiral system. We note that the continuum concepts of pressure and shear waves do not apply for waves in a lattice, especially in the case when the wavelength is comparable with the size of the elementary cell of the periodic structure. Special critical regimes are highlighted when vortex waveforms become dominant. Analytical findings are accompanied by illustrative numerical simulations.
\end{abstract}

\noindent{\it Keywords}: Chiral elastic lattice; lattice flux and circulation; vortex waveforms; wave polarisation.

\section{Introduction}
\label{Introduction}

Elastic lattices are relatively simple systems that exhibit many interesting dynamic properties, such as wave dispersion, filtering and dynamic anisotropy \cite{MartinssonMovchan2003,Ayzenberg-StepanenkoSlepyan2008,Colquitt2012}. Due to their discrete nature, lattice models allow the answer to fundamental questions on dynamic fracture problems, concerning in particular the analytical prediction of the speed of crack propagation and the explanation of crack tip instabilities \cite{MarderLiu1993,MarderGross1995,Slepyan2002,Slepyan2010,Nieves2016,Nieves2017}, that cannot be addressed by using continuum models. Homogenisation theories for discrete systems based on asymptotic techniques have been applied both in the static \cite{Bensoussan1978,BakhvalovPanasenko1984,ZhikovKozlovOleinik1994,Panasenko2005} and in the dynamic \cite{Craster2010,Antonakakis2014,MovchanSlepyan2014,Colquitt2015} regimes.

Polarisation of elastic waves in continuous media is well studied (see, for example, \cite{Love1892,Achenbach1973,Graff1975}). Recently, comparative analysis of polarisation of elastic waves in a continuum versus discrete medium has been performed in \cite{Carta2018}. It is well known that in a two-dimensional homogeneous infinite continuum two types of waves can propagate at different speeds, namely shear and pressure waves. In the former (or latter) case, the displacement vector is perpendicular (or parallel) to the wave vector. A triangular lattice approximates an isotropic continuum in the long wavelength limit or, equivalently, when the modulus of the wave vector tends to zero. For large values of the modulus of the wave vector, waves cannot be classified as shear or pressure waves. In \cite{Carta2018} two new quantities have been introduced, denoted as ``lattice flux'' and ``lattice circulation'', to characterise waves in the triangular lattice for any value of the wave vector. A decomposition of the displacement field has been proposed, whereby waves are described as a combination of flux-free and circulation-free components.

In this paper, we study a triangular lattice connected to a system of gyroscopic spinners. In this case, the trajectories of the lattice particles are not straight lines as in a classical triangular lattice, but ellipses. In some limit cases, discussed in depth in this work, the ellipses become circles. This special type of wave will be referred to as a ``vortex waveform''.

Throughout the present paper, we will refer to the triangular lattice connected to gyroscopic spinners as a ``chiral lattice''. According to the original definition by Lord Kelvin \cite{Kelvin1894}, an object is chiral if it cannot be superimposed onto its mirror image. The gyro-elastic lattice considered here is an ``active chiral'' medium, in which chirality is brought by the action of the gyroscopic spinners on the lattice particles. This type of chirality is different from the ``geometrical chirality'' discussed in \cite{Spadoni2009,SpaRuz2012,BacigalupoGambarotta2016,Tallarico2016} or from the interfacial wave guiding \cite{PalRuzzene2017}. Chirality discussed here can be used in unidirectional wave steering, as in \cite{Ni2015,SusstrunkHuber2015}, to create topological insulators.

The first model of an active chiral lattice was introduced in \cite{Brun2012}, where both a monatomic and a biatomic triangular lattice attached to a uniform system of gyroscopic spinners were studied. Furthermore, the homogenised equations of the discrete system were used to model a gyroscopic continuum, that was used to design a cloaking device. The monatomic gyro-elastic lattice proposed in \cite{Brun2012} was investigated in depth in \cite{Carta2014}, with special emphasis on tunable dynamic anisotropy and forced motions. Gyroscopic spinners were also employed to create localised waveforms in \cite{Carta2017} and in topological protection applications in \cite{Wang2015,Garau2018}. A hexagonal array of gyroscopes suspended by springs and magnetically coupled was built in \cite{Nash2015}, where unidirectional edge waves were experimentally observed.

Systems embedding gyroscopic spinners have many applications, especially in aerospace engineering \cite{Hughes1986,D'EleuterioHughes1987,Yamanaka1996,HassanpourHeppler2016a,HassanpourHeppler2016b}. For this reason, the theory of gyro-elastic continua has been developed in the literature (see, for example, \cite{D'EleuterioHughes1984,HughesD'Eleuterio1986}). Recently, attaching gyroscopic spinners to elastic beams in order to modify the dynamic properties of the beams has been proposed in \cite{Carta2018b,Nieves2018} and creating novel low-frequency resonators for seismic applications has been discussed in \cite{Carta2017b}.

The present paper is organised as follows. In Section \ref{EquationsDefinitionsSection}, the governing equations and the dispersion relation for a triangular lattice connected to a system of gyroscopic spinners are reviewed. In addition, the definitions of lattice flux and lattice circulation introduced in \cite{Carta2018} are discussed. In Section \ref{ChiralSection}, a decomposition of the displacement field in the chiral system is introduced. Moreover, lattice flux and lattice circulation are used to fully characterise waves propagating in the medium. The analysis is performed for the triangular chiral lattice studied in this paper; however, a similar formulation can be developed for any other type of gyro-elastic lattice, once the corresponding lattice flux and lattice circulation are derived. In Section \ref{ExamplesChiralSection}, the motion of the lattice for characteristic values of the wave vector is described. In particular, we show examples of vortex waveforms. In Section \ref{SectionLimitValuesAlpha}, the dynamic properties of the discrete system for limit values of the parameter characterising the spinners are investigated using asymptotic analyses. Finally, in Section \ref{Conclusions}, concluding remarks are provided.

\section{Governing equations and definitions}
\label{EquationsDefinitionsSection}

\begin{figure}%[!h]
\centering
\includegraphics[width=1.0\columnwidth]{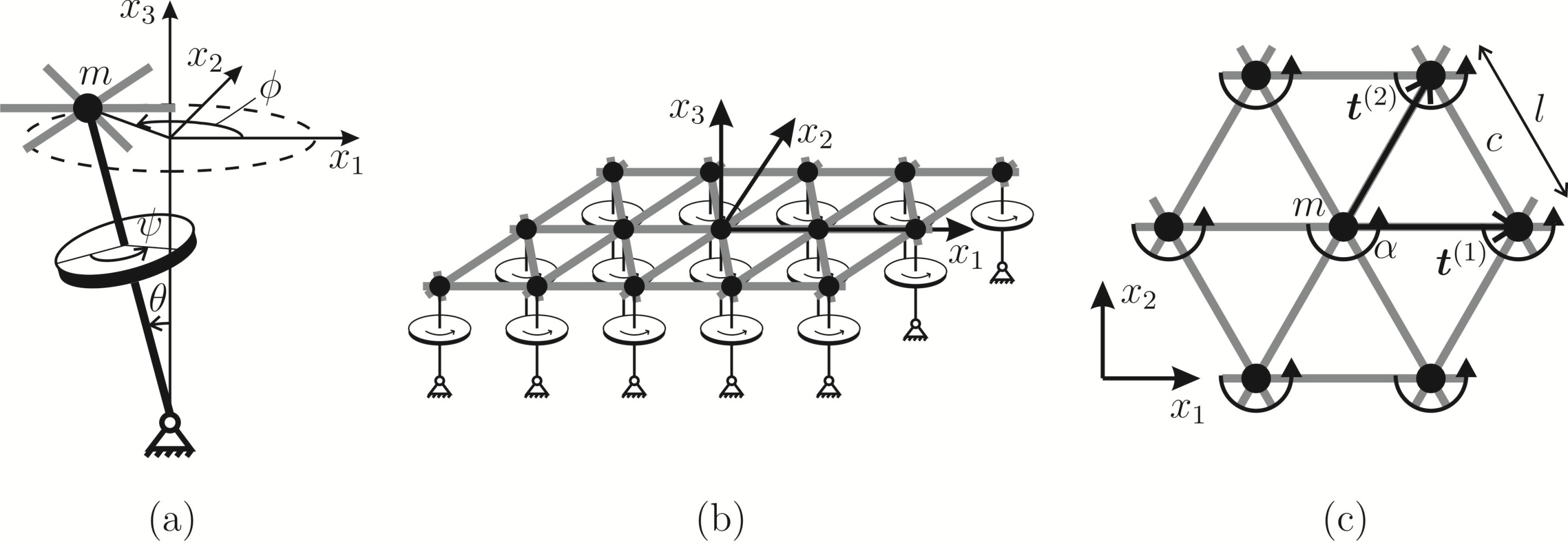}
\caption{\footnotesize (a) Representation of a gyroscopic spinner, where $\psi$, $\phi$ and $\theta$ are the angles of spin, precession and nutation respectively; (b) triangular elastic lattice connected to a system of gyroscopic spinners; (c) elementary cell of the lattice.}
\label{Lattice}
\end{figure}

We study an infinite, periodic triangular lattice of particles with mass $m$, connected by linear springs of stiffness $c$, length $l$ and negligible density. Each lattice particle is attached to a gyroscopic spinner (see Fig. \ref{Lattice}a), characterised by the spinner constant $\alpha$, which is a function of the geometry and spin rate of the spinner \cite{Brun2012}. The lattice is shown in Fig. \ref{Lattice}b and its elementary cell is presented in Fig. \ref{Lattice}c. We assume that the effect of gravity is negligible and the nutation angles $\theta$ of the spinners are small, so that the particles move in the $x_{1}x_{2}$-plane. This is the model system introduced in \cite{Brun2012,Carta2014}.

%%%%%%%%%%%%%%%%%%%%%%%%%%%%%%%%%%%%%%%%%%%%
%%%%%%%%%%%%%%%%%%%%%%%%%%%%%%%%%%%%%%%%%%%%

\subsection{Dispersion properties of the chiral lattice}
\label{DispersionSection}

In the time-harmonic regime, the displacement of a lattice particle $\bm{u}(\bm{x},t) = \bm{U} (\bm{x}) \mathrm{e}^{\mathrm{i} \omega t}$, where ${\bm x} = \left( x_1,x_2 \right)^{\mathrm{T}}$ is the position vector, $t$ is time and $\omega$ is the radian frequency. The displacements of the particles of the infinite periodic lattice are assumed to satisfy the Bloch-Floquet conditions:
\begin{equation}\label{BlochFloquetConditions}
{\bm U} \left( {\bm x} + n_1 {\bm t}^{(1)} + n_2 {\bm t}^{(2)} \right) = {\bm U} \left( {\bm x} \right) \mathrm{e}^{\mathrm{i} \, {\bm k} \cdot {\bm T} {\bm n}} \, .
\end{equation}
Here, ${\bm n} = \left( n_1,n_2 \right)^{\mathrm{T}}$ is the multi-index, ${\bm t}^{(1)}=\left( l , 0 \right)^{\mathrm{T}}$ and ${\bm t}^{(2)}=\left( l/2 , \sqrt{3} \, l/2 \right)^{\mathrm{T}}$ are the lattice vectors (see Fig. \ref{Lattice}c) and $\bm{k} = \left( k_{1},k_{2} \right)^{\textup{T}}$ is the wave vector. The matrix ${\bm T}$ is given by ${\bm T}=\left( {\bm t}^{(1)},{\bm t}^{(2)} \right)$.

The equations of motion of the chiral lattice can be written in the form \cite{Brun2012,Carta2014}
\begin{equation}\label{EquationsOfMotion}
\left[ \bm{C} - \omega^2\left( \bm{M}-\bm{A} \right) \right] \bm{U} = \bf{0} \, ,
\end{equation}
where $\bm{M} = \mathrm{diag}\left\{m,m\right\}$ is the mass matrix,
\begin{equation}\label{MatrixA}
\bm{A} =
\begin{pmatrix}
0 & -\mathrm{i} \alpha \\
\mathrm{i} \alpha & 0 \\
\end{pmatrix}
\end{equation}
is the spinner matrix and
\begin{equation}\label{matrixC}
{\bm C}=c
\begin{pmatrix}
3 - 2 \cos(\zeta l + \xi l) - \frac{1}{2} \left[\cos(\zeta l)+\cos{(\xi l)}\right] & \frac{\sqrt{3}}{2} \left[ \cos(\xi l)-\cos{(\zeta l)} \right] \\
\frac{\sqrt{3}}{2} \left[ \cos(\xi l)-\cos{(\zeta l)} \right] & 3 - \frac{3}{2} \left[ \cos(\zeta l)+\cos{(\xi l)} \right]
\end{pmatrix}
\end{equation}
is the stiffness matrix, where $\zeta = k_1/2 + \sqrt{3}k_2/2$ and $\xi = k_1/2 - \sqrt{3}k_2/2$.

We introduce the following normalisations:
\begin{equation}\label{NormalisedQuantities}
\begin{split}
&\tilde{\bm{x}} = \bm{x}/l \, , \; \tilde{\bm{U}} = \bm{U}/l \, , \; \tilde{\bm{u}} = \bm{u}/l \, , \; \tilde{\bm{T}} = \bm{T}/l \, , \; \tilde{\bm{k}} = \bm{k} \, l \, , \; \tilde{\zeta} = \zeta \, l \, , \; \tilde{\xi} = \xi \, l \, , \; \tilde{\bm{C}} = \bm{C}/c \, , \\
&\tilde{\bm{M}} = \bm{M}/m \, , \; \tilde{\bm{A}} = \bm{A}/m \, , \; \tilde{\alpha} = \alpha/m \, , \; \tilde{\omega} = \omega/\sqrt{c/m} \, , \; \tilde{t} = t \sqrt{c/m} \, , \\
\end{split}
\end{equation}
where the quantities with the symbol ``$\sim$'' are dimensionless.

The frequency $\tilde{\omega}$ and the wave vector $\tilde{\bm{k}}$ are related by the \emph{dispersion relation} of the system, which is given by \cite{Brun2012,Carta2014}
\begin{equation}\label{DispersionRelation}
\left( 1-\tilde{\alpha}^2 \right) \tilde{\omega}^4 - \mathrm{tr} (\tilde{{\bm C}}) \, \tilde{\omega}^2 +  \mathrm{det} (\tilde{{\bm C}}) = 0 \, .
\end{equation}
The two positive solutions of the above biquadratic equation in $\tilde{\omega}$ are
\begin{subequations}\label{Omega12}
\begin{align}
& \tilde{\omega}^{(1)}(\tilde{{\bm k}},\tilde{\alpha}) = \sqrt{\frac{\mathrm{tr}(\tilde{{\bm C}})-\sqrt{\mathrm{tr}^2(\tilde{{\bm C}})-4(1-\tilde{\alpha}^2)\mathrm{det}(\tilde{{\bm C}})}}{2(1-\tilde{\alpha}^2)}} \, , \\
& \tilde{\omega}^{(2)}(\tilde{{\bm k}},\tilde{\alpha}) = \sqrt{\frac{\mathrm{tr}(\tilde{{\bm C}})+\sqrt{\mathrm{tr}^2(\tilde{{\bm C}})-4(1-\tilde{\alpha}^2)\mathrm{det}(\tilde{{\bm C}})}}{2(1-\tilde{\alpha}^2)}} \, .
\end{align}
\end{subequations}
We note that $\tilde{\omega}^{(2)}$ takes imaginary values for $\tilde{\alpha} > 1$. After calculating the eigenfrequencies $\tilde{\omega}^{(j)}$ for a certain wave vector, the corresponding eigenvectors $\tilde{{\bm U}}^{(j)} = \tilde{{\bm U}}^{(j)}(\tilde{{\bm k}},\tilde{\alpha})$ ($j=1,2$) can be determined analytically from (\ref{EquationsOfMotion}).

\begin{figure}%[!h]
\centering
\includegraphics[width=1.0\columnwidth]{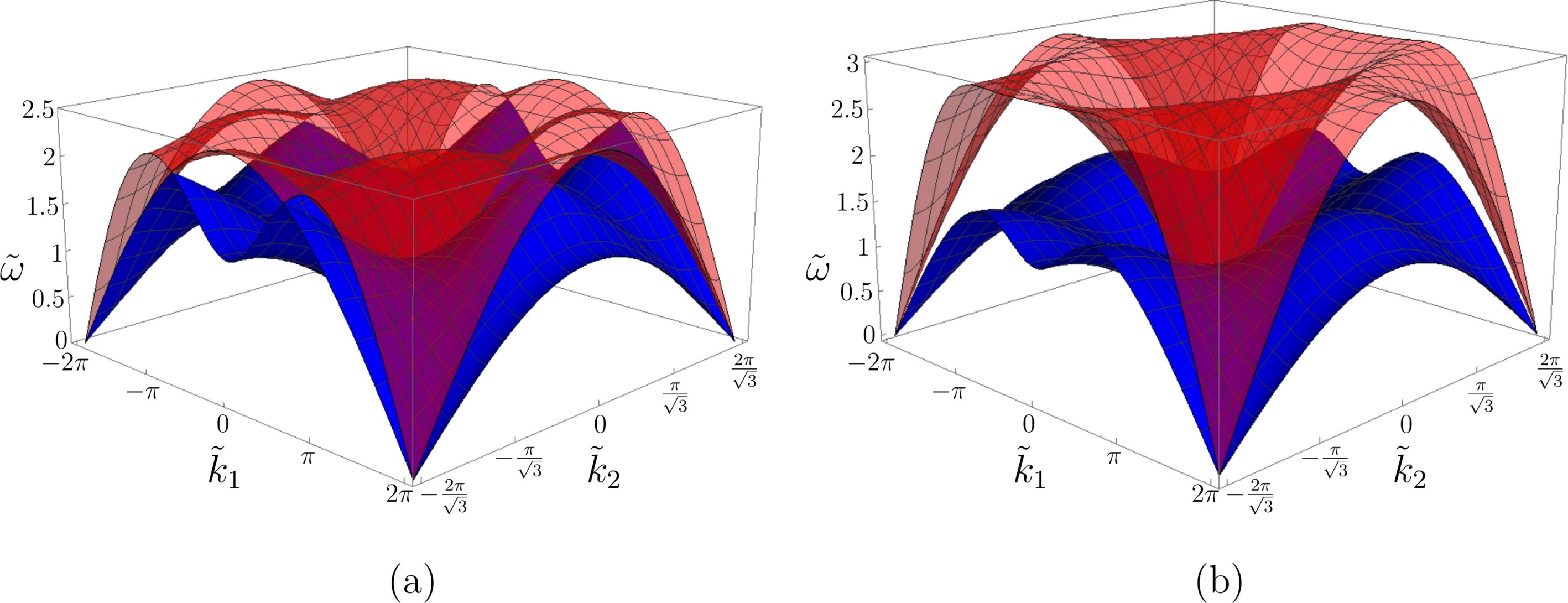}
\caption{\footnotesize Lower and upper dispersion surfaces for the triangular lattice in Fig. \ref{Lattice}, calculated for (a) $\tilde{\alpha} = 0$ and (b) $\tilde{\alpha} = 0.5$.}
\label{DispersionSurfaces}
\end{figure}

The dispersion surfaces for $\tilde{\alpha} = 0$ (non-chiral case) and $\tilde{\alpha} = 0.5$ are presented in Fig. \ref{DispersionSurfaces}. The main effect of the gyroscopic spinners on the dispersion surfaces of the lattice is to decrease (or increase) the values of $\tilde{\omega}^{(1)}$ (or $\tilde{\omega}^{(2)}$) for a fixed wave vector $\tilde{\bm{k}}$ \cite{Brun2012,Carta2014}. For $\tilde{\alpha} = 0$, the two dispersion surfaces touch at the Dirac points $\left( k_{1},k_{2} \right)^{\textup{T}} = \left( \pm 4 \pi/3,0 \right)^{\textup{T}}$ and $\left( k_{1},k_{2} \right)^{\textup{T}} = \left( \pm 2 \pi/3,\pm 2 \pi/\sqrt{3} \right)^{\textup{T}}$. For $\tilde{\alpha} > 0$, the dispersion surfaces no longer touch so that the Dirac cones are ``broken''.

%%%%%%%%%%%%%%%%%%%%%%%%%%%%%%%%%%%%%%%%%%%%
%%%%%%%%%%%%%%%%%%%%%%%%%%%%%%%%%%%%%%%%%%%%

\subsection{Definitions of lattice flux and lattice circulation}
\label{FluxCirculationSection}

As discussed in \cite{Carta2018} for the non-chiral case ($\tilde{\alpha} = 0$), waves propagating in a lattice can be characterised quantitatively by using the operators of \emph{lattice flux} and \emph{lattice circulation}. These are defined as (see \cite{Carta2018})
\begin{equation}\label{FluxLatticef}
\tilde{\Phi}_{\tilde{\bm{u}}} = \mathrm{i} \frac{\sqrt{3}}{2} \, \tilde{\bm{u}} \cdot \tilde{\bm{f}}
\end{equation}
and
\begin{equation}\label{CirculationLatticef}
\tilde{\Gamma}_{\tilde{\bm{u}}} = - \mathrm{i} \frac{\sqrt{3}}{2} \, \left( \tilde{\bm{u}} \times \tilde{\bm{f}} \right) \cdot \bm{e}_3 \, ,
\end{equation}
respectively, where $\bm{e}_3$ is the unit vector parallel to the $x_3$-axis. In (\ref{FluxLatticef}) and (\ref{CirculationLatticef}) we have also introduced the vector $\tilde{\bm{f}}$, given by
\begin{equation}\label{vectorf}
\tilde{\bm{f}} = \left( 2 \sin{(\tilde{\zeta}+\tilde{\xi})} + \sin{(\tilde{\zeta})} + \sin{(\tilde{\xi})} \, , \, \sqrt{3} \, \left[ \sin{(\tilde{\zeta})} - \sin{(\tilde{\xi})} \right] \right)^{\mathrm{T}} \, ,
\end{equation}
that depends on the geometry of the lattice.

In the long wavelength limit when $\left|\tilde{\bm{k}}\right| \to 0$, the lattice approximates a continuum. In this limit, $\tilde{\bm{f}} \sim 3 \tilde{\bm{k}}$. In a continuum, waves where the displacement $\tilde{\bm{u}}$ is perpendicular (or parallel) to the wave vector $\tilde{\bm{k}}$ are denoted as shear (or pressure) waves. Substituting $\tilde{\bm{f}} = 3 \tilde{\bm{k}}$ in (\ref{FluxLatticef}) and (\ref{CirculationLatticef}), we notice that in a continuum shear (or pressure) waves correspond to flux-free (or circulation-free) waves. For intermediate and large values of the modulus of the wave vector, the continuum concepts of shear and pressure waves cannot be applied to the lattice. Instead, we will employ the definitions (\ref{FluxLatticef}) and (\ref{CirculationLatticef}) to fully characterise waves propagating in the discrete system.

The vector $\tilde{\bm{u}}$ in (\ref{FluxLatticef}) and (\ref{CirculationLatticef}) represents the time-harmonic displacement of a lattice particle, calculated for a given eigenvector $\tilde{\bm{U}}$. For $\tilde{\alpha} < 1$, there are two eigenfrequencies and hence two eigenvectors $\tilde{\bm{U}}$ for any value of the wave vector $\tilde{\bm{k}}$. Denoting the coordinates of the central node of the lattice periodic cell shown in Fig. \ref{Lattice}c as $\tilde{\bm{x}}^0=(0,0)^{\mathrm{T}}$, the coordinates of the central node of the cell $\bm{n}$ ($\bm{n} \in \mathbb{Z}^2$) are $\tilde{\bm{x}} = \tilde{\bm{x}}^{(\bm{n},0)} = \tilde{\bm{x}}^0 + \tilde{\bm{T}}\bm{n}$. Using the Bloch-Floquet conditions (\ref{BlochFloquetConditions}), the time-harmonic displacement of a lattice particle for a given eigenvector is expressed by $\tilde{\bm{u}}^{(j)}(\tilde{\bm{x}},\tilde{t}) = \mathrm{Re}\left( \tilde{\bm{U}}^{(j)}(\tilde{\bm{x}}^0) \mathrm{e}^{\mathrm{i} \left( \tilde{\omega}^{(j)} \tilde{t} + \tilde{{\bm k}} \cdot \tilde{{\bm T}} {\bm n} \right)} \right)$ $(j=1,2)$, where $\tilde{\bm{U}}^{(j)}(\tilde{\bm{x}}^0) \mathrm{e}^{\mathrm{i} \left( \tilde{\omega}^{(j)} \tilde{t} + \tilde{{\bm k}} \cdot \tilde{{\bm T}} {\bm n} \right)} = \tilde{\bm{u}}^{(j)}(\tilde{\bm{x}}^0,\tilde{t}) \mathrm{e}^{\mathrm{i} \tilde{{\bm k}} \cdot \tilde{{\bm T}} {\bm n}}$ and $\tilde{\bm{u}}^{(j)}(\tilde{\bm{x}}^0,\tilde{t})$ is the displacement at $\tilde{\bm{x}}^0$. We now concentrate on the displacement $\tilde{\bm{u}}^{(j)}(\tilde{\bm{x}}^0,\tilde{t})$, that can also be written as
\begin{equation}\label{Displacements}
\tilde{\bm{u}}^{(j)} (\tilde{\bm{x}}^0,\tilde{t}) =
\begin{pmatrix}
\mathrm{Re}( \tilde{U}_1^{(j)} ) \cos(\tilde{\omega}^{(j)} \tilde{t}) - \mathrm{Im}( \tilde{U}_1^{(j)} ) \sin(\tilde{\omega}^{(j)} \tilde{t}) \\
\mathrm{Re}( \tilde{U}_2^{(j)} ) \cos(\tilde{\omega}^{(j)} \tilde{t}) - \mathrm{Im}( \tilde{U}_2^{(j)} ) \sin(\tilde{\omega}^{(j)} \tilde{t})
\end{pmatrix} \, ,
\quad j = 1,2 \, .
\end{equation}
The trajectory of the particle is an ellipse, since the eigenvectors are complex. In the non-chiral case ($\tilde{\alpha} = 0$) the particles trajectories are straight lines, since the eigenvectors are real. We also note that in the chiral lattice the eigenvectors are generally non-orthogonal. They satisfy the relation
\begin{equation}\label{InnerProductEigenvectors}
\left(\,{\overline{\tilde{\bm{U}}^{(j)}}}\,\right)^{\mathrm{T}} \, \tilde{\bm{U}}^{(i)} + \left(\,{\overline{\tilde{\bm{U}}^{(j)}}}\,\right)^{\mathrm{T}} \bm{R} \, \tilde{\bm{U}}^{(i)} = 0 \;\; \left( i \ne j \right)\text{,} \quad \text{with} \; \bm{R} =
\begin{pmatrix}
0 & \mathrm{i}\tilde{\alpha} \\
-\mathrm{i}\tilde{\alpha} & 0
\end{pmatrix}
 \, ,
\end{equation}
which reduces to the orthogonality condition when $\tilde{\alpha} = 0$.

\section{Wave characterisation in the chiral lattice}
\label{ChiralSection}

\begin{figure}%[!h]
\centering
\includegraphics[width=1.0\columnwidth]{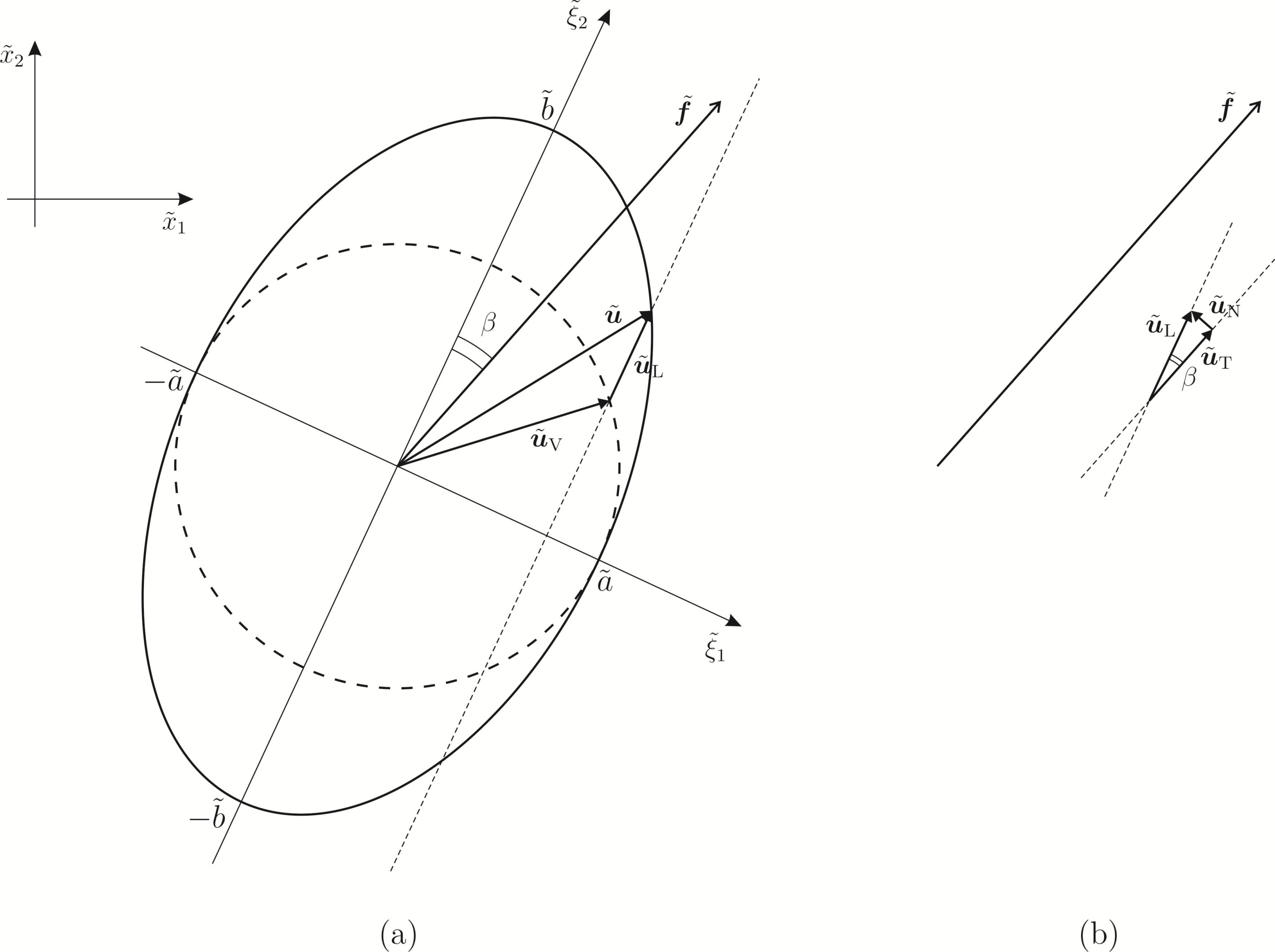}
\caption{\footnotesize (a) Generic elliptical trajectory of a lattice particle in the chiral lattice and decomposition of the displacement field $\tilde{\bm{u}}$ into a vortex component $\tilde{\bm{u}}_{\textup{V}}$ (corresponding to circular motion) and a straight-line component $\tilde{\bm{u}}_{\textup{L}}$ (corresponding to a straight-line motion parallel to the major axis of the ellipse); (b) a secondary decomposition of the straight-line field $\tilde{\bm{u}}_{\textup{L}}$ into a component $\tilde{\bm{u}}_{\textup{T}}$ tangential to $\tilde{\bm{f}}$ and a component $\tilde{\bm{u}}_{\textup{N}}$ normal to $\tilde{\bm{f}}$.}
\label{EllipticalTrajectoryChiral}
\end{figure}

As discussed in Section \ref{EquationsDefinitionsSection}\ref{FluxCirculationSection}, each particle in the chiral lattice describes an elliptical trajectory, shown in Fig. \ref{EllipticalTrajectoryChiral}a. The lengths of the minor and major semi-axes of the ellipse are denoted by $\tilde{a}$ and $\tilde{b}$, respectively. The angle between the major axis of the ellipse and the vector $\tilde{\bm{f}}$ is denoted by $\beta$. We note that $\beta$ is identical to the angle between the straight trajectory of a particle in the non-chiral case ($\tilde{\alpha} = 0$) and the vector $\tilde{\bm{f}}$. As $\tilde{\alpha}$ decreases, the resulting ellipses become narrower with the major axis direction remaining fixed. In the limit when $\tilde{\alpha} \to 0$ we retrieve the straight-line motion (in the direction of the major axis) we observe in the corresponding non-chiral case ($\tilde{\alpha} = 0$).

%%%%%%%%%%%%%%%%%%%%%%%%%%%%%%%%%%%%%%%%%%%%
%%%%%%%%%%%%%%%%%%%%%%%%%%%%%%%%%%%%%%%%%%%%

\subsection{Decomposition of the displacement field}
\label{DecompositionSection}

As shown in Fig. \ref{EllipticalTrajectoryChiral}a, the displacement $\tilde{\bm{u}}$ can be decomposed into a component parallel to the major axis of the ellipse, denoted as $\tilde{\bm{u}}_{\textup{L}}$, and a component $\tilde{\bm{u}}_{\textup{V}}$ whose end describes a circular trajectory. The subscript ``L'' in $\tilde{\bm{u}}_{\textup{L}}$ stands for line, since a particle having that displacement would move in a straight line parallel to the major axis. The subscript ``V'' in $\tilde{\bm{u}}_{\textup{V}}$ stands for vortex, since it corresponds to a circular trajectory.

The straight-line component $\tilde{\bm{u}}_{\textup{L}}$ can be further decomposed into a component $\tilde{\bm{u}}_{\textup{T}}$ parallel to $\tilde{\bm{f}}$, characterised by zero circulation, and a component $\tilde{\bm{u}}_{\textup{N}}$ perpendicular to $\tilde{\bm{f}}$, having zero flux (see Fig. \ref{EllipticalTrajectoryChiral}b). This secondary decomposition was also used in \cite{Carta2018} to characterise waves in the non-chiral case ($\tilde{\alpha} = 0$), where the displacement $\tilde{\bm{u}}=\tilde{\bm{u}}_{\textup{L}}$ ($\tilde{\bm{u}}_{\textup{V}}={\bf 0}$). Therefore, the displacement field in the chiral lattice $\tilde{\bm{u}}=\tilde{\bm{u}}_{\textup{L}}+\tilde{\bm{u}}_{\textup{V}}$ consists of a ``non-chiral'' component $\tilde{\bm{u}}_{\textup{L}}$ and a ``chiral'' component $\tilde{\bm{u}}_{\textup{V}}$. Both $\tilde{\bm{u}}_{\textup{L}}$ and $\tilde{\bm{u}}_{\textup{V}}$ are functions of the wave number $\tilde{\bm{k}}$ and the spinner constant $\tilde{\alpha}$.

It is important to note that the decomposition of the displacement field is not unique and the decomposition introduced above emphasises and distils the circular (or vortex) displacement field, associated with chiral motion. One such alternative is to decompose the displacement field into components perpendicular and parallel to the major axis of the elliptical, chiral displacement. In this alternative decomposition, however, the two components would include the vortex motion, and a comparison with the non-chiral case ($\tilde{\alpha} = 0$) would be less straightforward.

The degree of chirality in a lattice with gyroscopic spinners can be measured by the following parameter:
\begin{equation}\label{ChiralityFactor}
\chi^{(j)} = \frac{\tilde{a}^{(j)}}{\tilde{b}^{(j)}} \, , \quad j=1,2 \, ,
\end{equation}
which represents the ratio of the length of the minor semi-axis to the length of the major semi-axis of the ellipse. We note that $0 \le \chi^{(j)} \le 1$, where the lower limit $\chi^{(j)}=0$ is found in the non-chiral case ($\tilde{\alpha} = 0$) where the trajectory is always a straight line, while the upper limit $\chi^{(j)}=1$ is reached in the chiral lattice in some special cases when the trajectory is a circle (vortex waveforms).

The lengths of the minor and major semi-axes can be determined from the eigenvectors calculated from the dispersive properties of the system. Using (\ref{Displacements}), the canonical equation for the ellipse can be written as
\begin{equation}\label{EquationEllipse}
\frac{1}{\mathrm{det}(\bm{B}^{(j)})} \left[ \bm{B}_{11}^{(j)} \tilde{x}_1^2 - 2 \bm{B}_{12}^{(j)} \tilde{x}_1 \tilde{x}_2 + \bm{B}_{22}^{(j)} \tilde{x}_2^2 \right] = 1 \, ,
\end{equation}
where the components of the matrix $\bm{B}^{(j)}=\left(\bm{B}^{(j)}\right)^{\mathrm{T}}$ ($j=1,2$) are
\begin{subequations}\label{MatrixB}
\begin{alignat}{3}
\bm{B}_{11}^{(j)} & = \mathrm{Re}\left(\tilde{U}_2^{(j)}\right)^2 + \mathrm{Im}\left(\tilde{U}_2^{(j)}\right)^2 \, , \\
\bm{B}_{12}^{(j)} & = \bm{B}_{21}^{(j)} = - \left[ \mathrm{Re}\left(\tilde{U}_1^{(j)}\right)\mathrm{Re}\left(\tilde{U}_2^{(j)}\right)+\mathrm{Im}\left(\tilde{U}_1^{(j)}\right)\mathrm{Im}\left(\tilde{U}_2^{(j)}\right) \right] \, , \\
\bm{B}_{22}^{(j)} & = \mathrm{Re}\left(\tilde{U}_1^{(j)}\right)^2 + \mathrm{Im}\left(\tilde{U}_1^{(j)}\right)^2 \, .
\end{alignat}
\end{subequations}
The eigenvalues of $\bm{B}^{(j)}$ are given by
\begin{equation}\label{EigenvaluesEllipse}
\lambda_{\pm}^{(j)} = \frac{\mathrm{tr}\left(\bm{B}^{(j)}\right) \pm \sqrt{\mathrm{tr}^2\left(\bm{B}^{(j)}\right) - 4 \mathrm{det}\left(\bm{B}^{(j)}\right)}}{2 \mathrm{det}\left(\bm{B}^{(j)}\right)} \, ,
\end{equation}
while the eigenvectors of $\bm{B}^{(j)}$ are expressed by
\begin{equation}\label{EigenvectorsEllipse}
\bm{V}_{\pm}^{(j)} =
\begin{pmatrix}
\frac{\bm{B}_{11}^{(j)} - \bm{B}_{22}^{(j)} \pm \sqrt{\mathrm{tr}^2\left(\bm{B}^{(j)}\right) - 4 \mathrm{det}\left(\bm{B}^{(j)}\right)}}{-2 \bm{B}_{12}^{(j)}} \\
1
\end{pmatrix} \, .
\end{equation}

The lengths of the minor and major semi-axes of the ellipse are then given by
\begin{equation}\label{abEllipse}
\tilde{a}^{(j)} = \frac{1}{\sqrt{\lambda_{+}^{(j)}}} \; , \quad \tilde{b}^{(j)} = \frac{1}{\sqrt{\lambda_{-}^{(j)}}} \, .
\end{equation}
The direction of the major axis is defined by $\bm{V}_{-}^{(j)}$. The angle $\beta^{(j)}$ ($j=1,2$) is the angle between the major axis of the ellipse and the vector $\tilde{\bm{f}}$. In this paper, we take $\beta^{(j)}$ as the acute angle between $\bm{V}_{-}^{(j)}$ and $\tilde{\bm{f}}$, such that $0 \le \beta^{(j)} \le \pi/2$:
\begin{subequations}\label{FormulaBeta}
\begin{align}
& \beta^{(j)} = \arccos{\left( \frac{\bm{V}_{-}^{(j)} \cdot \tilde{\bm{f}}}{\Big|\tilde{\bm{f}}\Big| \Big|\tilde{\bm{V}_{-}^{(j)}}\Big|} \right)} \quad \text{if } \bm{V}_{-}^{(j)} \cdot \tilde{\bm{f}}>0 \, , \\
& \beta^{(j)} = \pi - \arccos{\left( \frac{\bm{V}_{-}^{(j)} \cdot \tilde{\bm{f}}}{\Big|\tilde{\bm{f}}\Big| \Big|\tilde{\bm{V}_{-}^{(j)}}\Big|} \right)} \quad \text{if } \bm{V}_{-}^{(j)} \cdot \tilde{\bm{f}}<0 \, .
\end{align}
\end{subequations}

\subsection{Flux and circulation in the chiral lattice}
\label{FluxCirculationSectionChiral}

As shown in (\ref{FluxLatticef}) and (\ref{CirculationLatticef}), the flux and circulation are pure imaginary quantities, with moduli $\left| \tilde{\Phi}_{\tilde{\bm{u}}} \right|$ and $\left| \tilde{\Gamma}_{\tilde{\bm{u}}} \right|$ respectively. In this paper, we denote by $\Arrowvert\tilde{\Phi}_{\tilde{\bm{u}}}\Arrowvert = \max\left| \tilde{\Phi}_{\tilde{\bm{u}}} \right|$ and $\Arrowvert\tilde{\Gamma}_{\tilde{\bm{u}}}\Arrowvert = \max\left| \tilde{\Gamma}_{\tilde{\bm{u}}} \right|$ the ``amplitudes'' of flux and circulation, respectively.

As discussed in \cite{Carta2018}, in the non-chiral case ($\tilde{\alpha} = 0$) the flux amplitude for the lower dispersion surface $\Arrowvert\tilde{\Phi}_{\tilde{\bm{u}}}^{(1)}\Arrowvert$ is equal to the circulation amplitude for the upper dispersion surface $\Arrowvert\tilde{\Gamma}_{\tilde{\bm{u}}}^{(2)}\Arrowvert$, if the eigenvectors corresponding to the two dispersion surfaces are normalised to have the same modulus; furthermore, $\Arrowvert\tilde{\Phi}_{\tilde{\bm{u}}}^{(2)}\Arrowvert = \Arrowvert\tilde{\Gamma}_{\tilde{\bm{u}}}^{(1)}\Arrowvert$. This is due to the orthogonality of the eigenvectors in the non-chiral case (see (\ref{InnerProductEigenvectors}) for $\tilde{\alpha}=0$), as discussed in \cite{Carta2018}. In the chiral lattice generally these relations do not hold, since the eigenvectors are not orthogonal.

\begin{figure}%[!h]
\centering
\includegraphics[width=1.0\columnwidth]{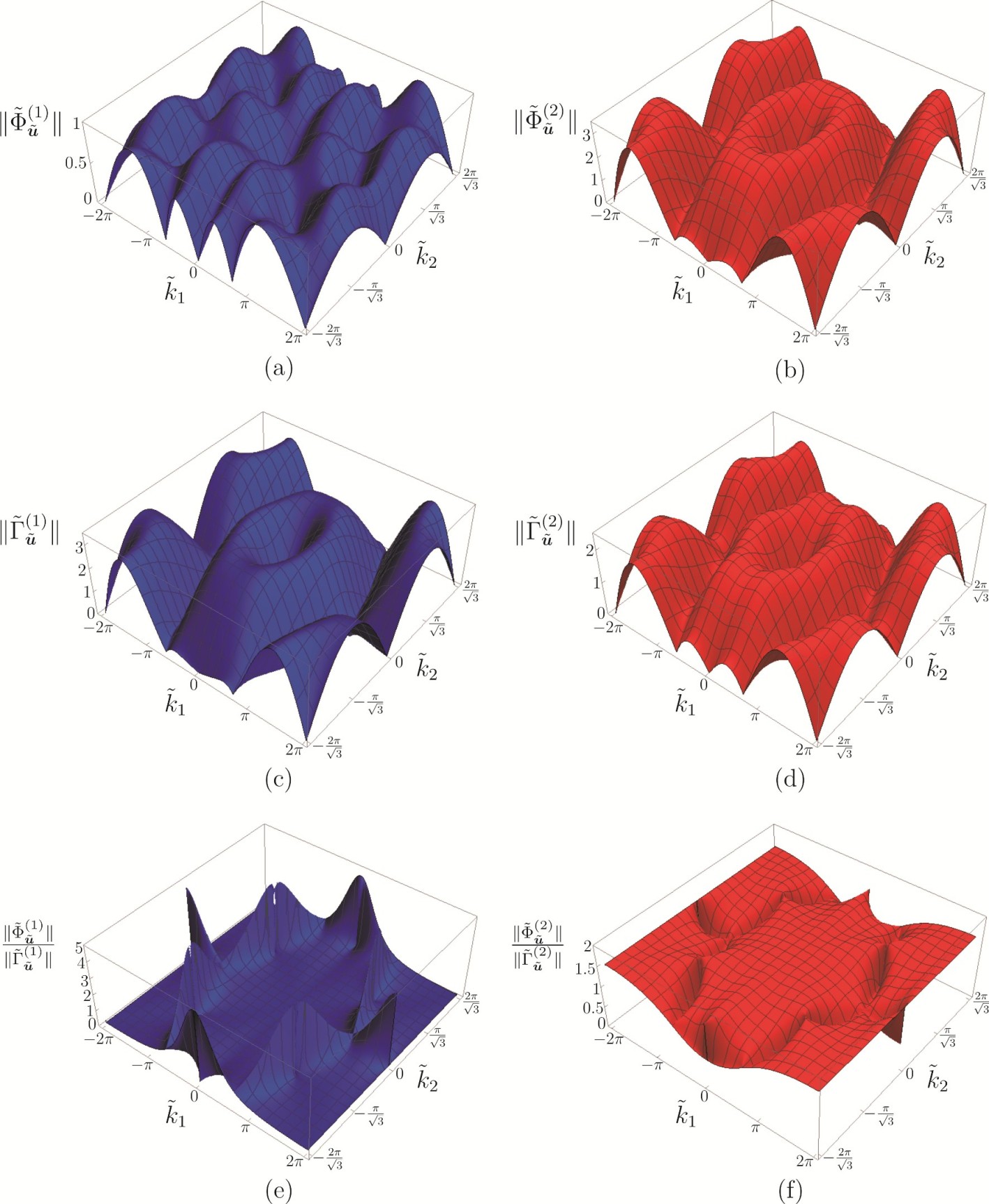}
\caption{\footnotesize The amplitudes (a) $\Arrowvert\tilde{\Phi}_{\tilde{\bm{u}}}^{(1)}\Arrowvert$, (b) $\Arrowvert\tilde{\Phi}_{\tilde{\bm{u}}}^{(2)}\Arrowvert$, (c) $\Arrowvert\tilde{\Gamma}_{\tilde{\bm{u}}}^{(1)}\Arrowvert$, (d) $\Arrowvert\tilde{\Gamma}_{\tilde{\bm{u}}}^{(2)}\Arrowvert$, and the ratios (e) $\Arrowvert\tilde{\Phi}_{\tilde{\bm{u}}}^{(1)}\Arrowvert/\Arrowvert\tilde{\Gamma}_{\tilde{\bm{u}}}^{(1)}\Arrowvert$, (f) $\Arrowvert\tilde{\Phi}_{\tilde{\bm{u}}}^{(2)}\Arrowvert/\Arrowvert\tilde{\Gamma}_{\tilde{\bm{u}}}^{(2)}\Arrowvert$ in the $\tilde{\bm{k}}$-plane, calculated for the chiral lattice with $\tilde{\alpha}=0.5$.}
\label{3DquantitiesChiral}
\end{figure}

The three-dimensional representations in the $\tilde{\bm{k}}$-plane of the amplitudes of flux and circulation for both dispersion surfaces are plotted in Figs. \ref{3DquantitiesChiral}a-\ref{3DquantitiesChiral}d for a representative value of the spinner constant $\tilde{\alpha}=0.5$. The qualitative features shown in Figs. \ref{3DquantitiesChiral}a-\ref{3DquantitiesChiral}d persist for all values of $\tilde{\alpha}$ for the lower dispersion surface and for $0<\tilde{\alpha}<1$ for the upper dispersion surface. The angles $\beta^{(1)}$ and $\beta^{(2)}$ are the same as those found in the non-chiral case (see Figs. 5c and 5d in \cite{Carta2018}).

We point out that the maps of flux and circulation in the $\tilde{\bm{k}}$-plane depend on the chosen normalisation of the eigenvectors. In the computations presented in this paper, the eigenvectors in (\ref{Displacements}) are normalised such that $\tilde{b} = 1$. This is in agreement with the normalisation adopted in \cite{Carta2018} for the non-chiral case ($\tilde{\alpha} = 0$), whereby the maximum straight-line displacement of each lattice particle is $1$. However, it is important to note that the ratio of flux to circulation for each dispersion surface is independent of the normalisation of the eigenvectors. The ratio gives a measure, independent of the normalisation of the eigenvectors, of the relative contributions of flux and circulation for a given wave. In Figs. \ref{3DquantitiesChiral}e and \ref{3DquantitiesChiral}f we show the three-dimensional representations of the ratios $\Arrowvert\tilde{\Phi}_{\tilde{\bm{u}}}^{(1)}\Arrowvert/\Arrowvert\tilde{\Gamma}_{\tilde{\bm{u}}}^{(1)}\Arrowvert$ and $\Arrowvert\tilde{\Phi}_{\tilde{\bm{u}}}^{(2)}\Arrowvert/\Arrowvert\tilde{\Gamma}_{\tilde{\bm{u}}}^{(2)}\Arrowvert$, respectively. In particular, we observe that in the long wavelength limit circulation (or flux) is dominant on the lower (or upper) surface.

%%%%%%%%%%%%%%%%%%%%%%%%%%%%%%%%%%%%%%%%%%%%
%%%%%%%%%%%%%%%%%%%%%%%%%%%%%%%%%%%%%%%%%%%%

From Figs. \ref{3DquantitiesChiral}a-\ref{3DquantitiesChiral}d it can be noted that the amplitudes of flux and circulation are continuous functions of $\tilde{\bm{k}}$. From the figures, it is also apparent that there are no lines where either the flux or the circulation are zero, while in the non-chiral case ($\tilde{\alpha}=0$), as observed in \cite{Carta2018}, there are special lines in the $\tilde{\bm{k}}$-plane where waves are either flux-free or circulation-free, even for large values of $\left|\tilde{\bm{k}}\right|$.

%%%%%%%%%%%%%%%%%%%%%%%%%%%%%%%%%%%%%%%%%%%%
\begin{figure}%[!h]
\centering
\includegraphics[width=1.0\columnwidth]{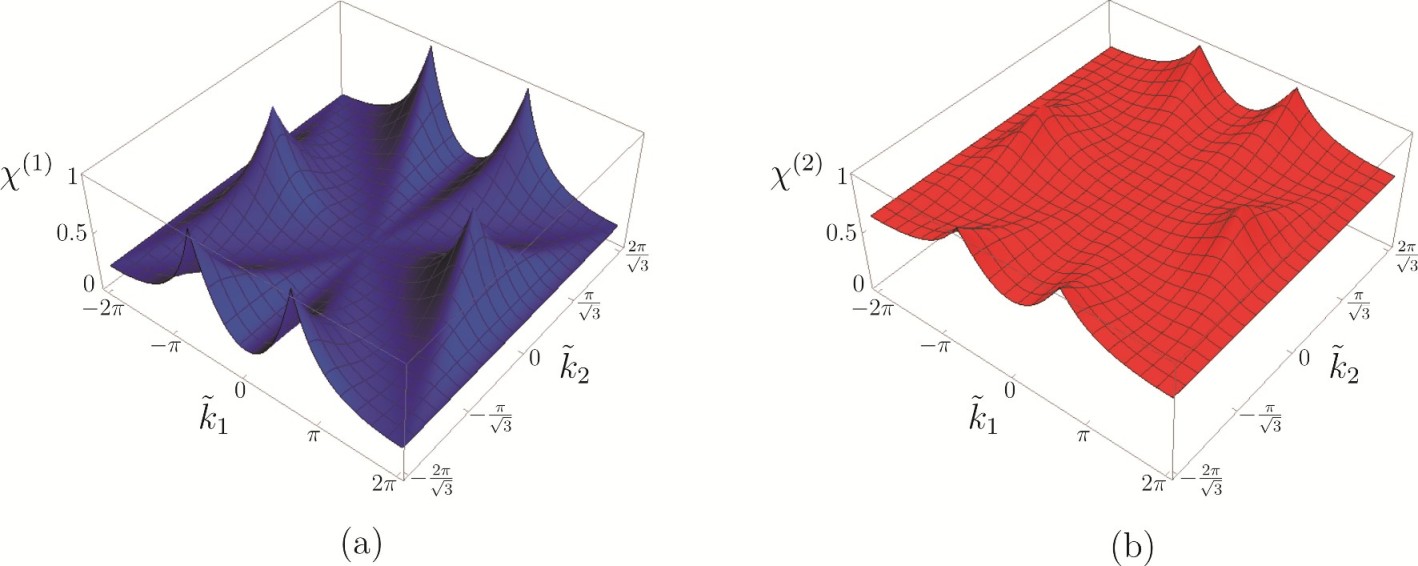}
\caption{\footnotesize Ratios (a) $\chi^{(1)}$ and (b) $\chi^{(2)}$ in the $\tilde{\bm{k}}$-plane, calculated for the chiral lattice with $\tilde{\alpha}=0.5$.}
\label{ChijChiral}
\end{figure}
%%%%%%%%%%%%%%%%%%%%%%%%%%%%%%%%%%%%%%%%%%%%

In Figs. \ref{ChijChiral}a and \ref{ChijChiral}b, we show $\chi^{(1)}$ and $\chi^{(2)}$ as functions of the wave vector. Interestingly, $\chi^{(2)} \ge \chi^{(1)}$ for any value of the wave vector. In addition, from Figs. \ref{ChijChiral}a and \ref{ChijChiral}b, we observe that $\chi^{(2)} = \chi^{(1)} = 1$ at all the points D in Fig. \ref{StationaryPoints}. At these points, every particle in the chiral lattice moves in a circle. In the non-chiral case ($\tilde{\alpha} = 0$), points D are vertices of Dirac cones.

\begin{figure}%[!h]
\centering
\includegraphics[width=1.0\columnwidth]{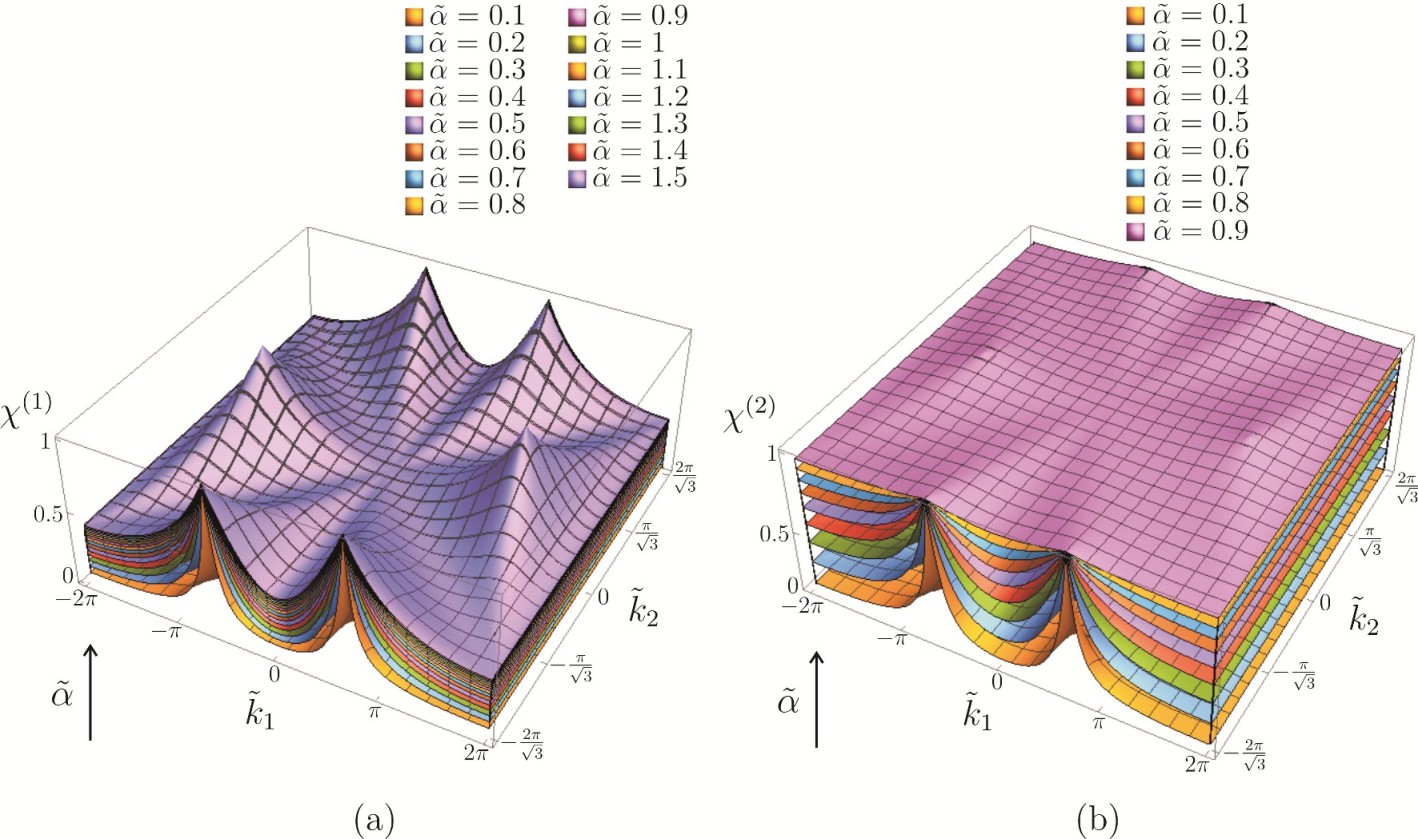}
\caption{\footnotesize (a) $\chi^{(1)}$ and (b) $\chi^{(2)}$ as functions of $\tilde{\bm{k}}$, determined for different values of $\tilde{\alpha}$, as indicated in the legends.}
\label{ChiVsAlpha}
\end{figure}

Figs. \ref{ChiVsAlpha}a and \ref{ChiVsAlpha}b show $\chi^{(1)}$ and $\chi^{(2)}$ in the $\tilde{\bm{k}}$-plane for different values of $\tilde{\alpha}$. We notice that both $\chi^{(1)}$ and $\chi^{(2)}$ increase with the spinner constant for any value of the wave vector. We have checked this analytically by verifying that $\partial \chi^{(j)}/\partial \tilde{\alpha} > 0$ ($j=1,2$) for any $\tilde{\bm{k}}$ and for any $\tilde{\alpha}$ (the results are not included here for brevity).

%%%%%%%%%%%%%%%%%%%%%%%%%%%%%%%%%%%%%%%%%%%%
%%%%%%%%%%%%%%%%%%%%%%%%%%%%%%%%%%%%%%%%%%%%

The flux and circulation of the total displacement field $\tilde{\bm{u}}$ can be decomposed into two components, one associated with the vortex field $\tilde{\bm{u}}_{\textup{V}}$ and the other with the straight-line field $\tilde{\bm{u}}_{\textup{L}}$. The straight-line displacement can be further decomposed into a circulation-free and a flux-free component. Referring to Fig. \ref{EllipticalTrajectoryChiral}a, the total displacement can be written in the rotated frame aligned with the principal axes $(\tilde{\xi}_1,\tilde{\xi}_2)$ of the ellipse as
\begin{equation}\label{DecompositionVortexLinear}
\tilde{\bm{u}} = \tilde{\bm{u}}_{\textup{V}}+\tilde{\bm{u}}_{\textup{L}} =
\begin{pmatrix}
\tilde{a} \cos{(\tilde{\omega}\tilde{t})} \\
\tilde{a} \sin{(\tilde{\omega}\tilde{t})}
\end{pmatrix}
+
\begin{pmatrix}
0 \\
(\tilde{b}-\tilde{a}) \sin{(\tilde{\omega}\tilde{t})}
\end{pmatrix}
\, .
\end{equation}
Using (\ref{FluxLatticef}) and (\ref{CirculationLatticef}), we find that the flux and circulation associated with the vortex field are given by
\begin{equation}\label{FluxCirculationVortex}
\tilde{\Phi}_{\tilde{\bm{u}}_{\textup{V}}} = \mathrm{i} \frac{\sqrt{3}}{2} \tilde{a} \left| \tilde{\bm{f}} \right| \sin{(\tilde{\omega}\tilde{t}+\beta)} \quad \mathrm{and} \quad \tilde{\Gamma}_{\tilde{\bm{u}}_{\textup{V}}} = \mathrm{i} \frac{\sqrt{3}}{2} \tilde{a} \left| \tilde{\bm{f}} \right| \cos{(\tilde{\omega}\tilde{t}+\beta)} \, ,
\end{equation}
respectively. Accordingly, the flux and circulation of the vortex field differ in phase by $\pi/2$ and have the same amplitude, namely
\begin{equation}\label{FluxCirculationVortexAmplitudes}
\Arrowvert\tilde{\Phi}_{\tilde{\bm{u}}_{\textup{V}}}\Arrowvert = \Arrowvert\tilde{\Gamma}_{\tilde{\bm{u}}_{\textup{V}}}\Arrowvert = \frac{\sqrt{3}}{2} \tilde{a} \left| \tilde{\bm{f}} \right| \, .
\end{equation}
Such a vortex field possesses the following properties:
\begin{itemize}
    \item the trajectories of nodal points within the lattice are circular, with a phase shift present between different elementary cells;
    \item the maximum amplitudes of lattice flux and lattice circulation are equal.
\end{itemize}
This is a third fundamental field present in characterising waves in chiral lattices, in addition to the flux-free and circulation-free fields observed in non-chiral case ($\tilde{\alpha} = 0$), as discussed in \cite{Carta2018}.

The straight-line displacement field can be decomposed into a component tangential to $\tilde{\bm{f}}$ and a component normal to $\tilde{\bm{f}}$, such that $\tilde{\bm{u}}_{\textup{L}} = \tilde{\bm{u}}_{\textup{T}}+\tilde{\bm{u}}_{\textup{N}}$ (see Fig. \ref{EllipticalTrajectoryChiral}b). The tangential component $\tilde{\bm{u}}_{\textup{T}}$ has zero circulation, while its flux is equal to
\begin{equation}\label{FluxTangential}
\tilde{\Phi}_{\tilde{\bm{u}}_{\textup{T}}} = \mathrm{i} \frac{\sqrt{3}}{2} \left(\tilde{b}-\tilde{a}\right) \left| \tilde{\bm{f}} \right| \cos{(\beta)} \sin{(\tilde{\omega}\tilde{t})} \, ,
\end{equation}
with amplitude
\begin{equation}\label{FluxTangentialAmplitude}
\Arrowvert\tilde{\Phi}_{\tilde{\bm{u}}_{\textup{T}}}\Arrowvert = \frac{\sqrt{3}}{2} \left(\tilde{b}-\tilde{a}\right) \left| \tilde{\bm{f}} \right| \cos{(\beta)} \, .
\end{equation}
On the other hand, the normal component $\tilde{\bm{u}}_{\textup{N}}$ is characterised by zero flux and non-zero circulation, given by
\begin{equation}\label{CirculationNormal}
\tilde{\Gamma}_{\tilde{\bm{u}}_{\textup{N}}} = - \mathrm{i} \frac{\sqrt{3}}{2} \left(\tilde{b}-\tilde{a}\right) \left| \tilde{\bm{f}} \right| \sin{(\beta)} \sin{(\tilde{\omega}\tilde{t})} \, ,
\end{equation}
having amplitude
\begin{equation}\label{CirculationNormalAmplitude}
\Arrowvert\tilde{\Gamma}_{\tilde{\bm{u}}_{\textup{N}}}\Arrowvert = \frac{\sqrt{3}}{2} \left(\tilde{b}-\tilde{a}\right) \left| \tilde{\bm{f}} \right| \sin{(\beta)} \, .
\end{equation}

\begin{figure}%[!h]
\centering
\includegraphics[width=1.0\columnwidth]{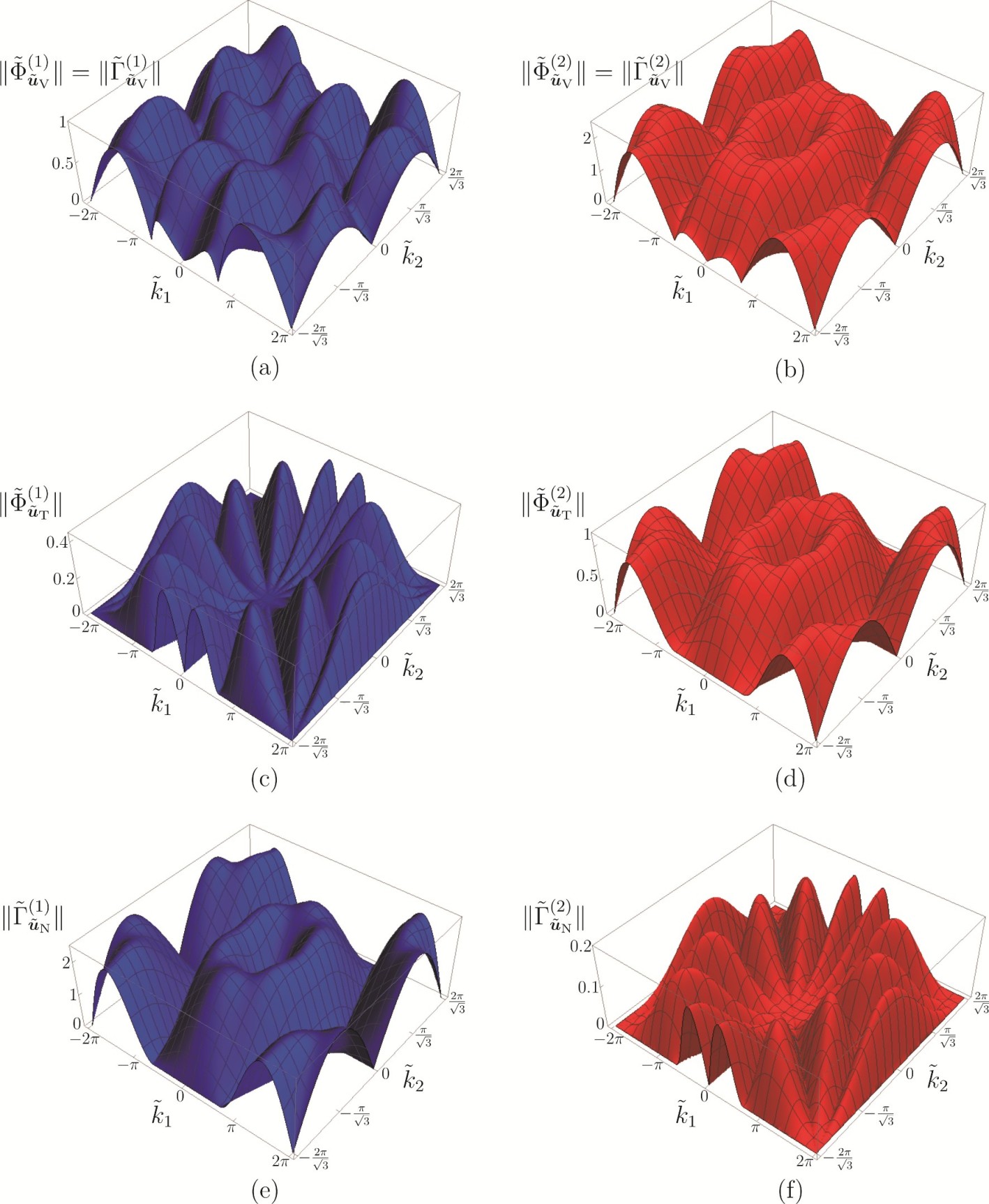}
\caption{\footnotesize Non-zero amplitudes of flux and circulation associated with (a,b) the vortex component $\tilde{\bm{u}}_{\textup{V}}$, (c,d) straight-line tangential component $\tilde{\bm{u}}_{\textup{T}}$ and (e,f) straight-line normal component $\tilde{\bm{u}}_{\textup{N}}$ in the $\tilde{\bm{k}}$-plane. The value of the spinner constant $\tilde{\alpha} = 0.5$ is the same as in Fig. \ref{3DquantitiesChiral}.}
\label{3DquantitiesComponentsChiral}
\end{figure}

The amplitudes of flux and circulation for the displacement components $\tilde{\bm{u}}_{\textup{V}}$, $\tilde{\bm{u}}_{\textup{T}}$ and $\tilde{\bm{u}}_{\textup{N}}$ as functions of the wave vector $\tilde{\bm{k}}$ are presented in Fig. \ref{3DquantitiesComponentsChiral}. The same normalisation of the eigenvectors as for the diagrams in Fig. \ref{3DquantitiesChiral} has been used, namely $\tilde{b}^{(j)} = 1$ ($j=1,2$).

%%%%%%%%%%%%%%%%%%%%%%%%%%%%%%%%%%%%%%%%%%%%
%%%%%%%%%%%%%%%%%%%%%%%%%%%%%%%%%%%%%%%%%%%%

Comparing Figs. \ref{3DquantitiesComponentsChiral}a and \ref{3DquantitiesComponentsChiral}b, we note that the contribution of the vortex component to the total displacement is larger for the upper surface. This in agreement with the diagrams in Figs. \ref{ChijChiral}a and \ref{ChijChiral}b, whereby $\chi^{(2)} \ge \chi^{(1)}$ and hence the radius of the circular trajectory for the upper surface is larger than that for the lower surface ($\tilde{a}^{(2)} \ge \tilde{a}^{(1)}$) keeping the length of the major semi-axis the same ($\tilde{b}^{(2)} = \tilde{b}^{(1)} = 1$). Concerning the straight-line component of the displacement, Figs. \ref{3DquantitiesComponentsChiral}c-\ref{3DquantitiesComponentsChiral}f reveal that the flux and circulation in the chiral lattice have features similar to those identified in the non-chiral case ($\tilde{\alpha} = 0$) \cite{Carta2018}. In particular, $\Arrowvert\tilde{\Phi}_{\tilde{\bm{u}}_{\textup{T}}}^{(1)}\Arrowvert=\Arrowvert\tilde{\Gamma}_{\tilde{\bm{u}}_{\textup{N}}}^{(2)}\Arrowvert=0$ in the lines given by $\arctan{\left( \tilde{k}_2/\tilde{k}_1 \right)} = (n-1)\pi/6$ with $n=1,..,12$, while $\Arrowvert\tilde{\Gamma}_{\tilde{\bm{u}}_{\textup{N}}}^{(1)}\Arrowvert=\Arrowvert\tilde{\Phi}_{\tilde{\bm{u}}_{\textup{T}}}^{(2)}\Arrowvert=0$ in the hexagon connecting the points D in Fig. \ref{StationaryPoints}. Additionally, we note that the ratio of the flux to the circulation for the lower surface ($\Arrowvert\tilde{\Phi}_{\tilde{\bm{u}}_{\textup{T}}}^{(1)}\Arrowvert/\Arrowvert\tilde{\Gamma}_{\tilde{\bm{u}}_{\textup{N}}}^{(1)}\Arrowvert$) is generally smaller than that for the upper surface (($\Arrowvert\tilde{\Phi}_{\tilde{\bm{u}}_{\textup{T}}}^{(2)}\Arrowvert/\Arrowvert\tilde{\Gamma}_{\tilde{\bm{u}}_{\textup{N}}}^{(2)}\Arrowvert$)). This means that for $\tilde{\alpha} = 0.5$ the straight-line component of the displacement is of flux-free type for the lower surface and of circulation-free type for the upper surface. However, differently from the non-chiral case ($\tilde{\alpha} = 0$), here the contribution of the vortex component (characterised by equal amplitudes of flux and circulation) is significant in that it changes the overall motion of the lattice.

The amplitudes of flux and circulation of the total displacement $\tilde{\bm{u}}$, shown in Fig. \ref{3DquantitiesChiral}, can be obtained in terms of the amplitudes of flux and circulation of the displacement components $\tilde{\bm{u}}_{\textup{V}}$, $\tilde{\bm{u}}_{\textup{T}}$ and $\tilde{\bm{u}}_{\textup{N}}$, presented in Fig. \ref{3DquantitiesComponentsChiral}, as follows:
\begin{subequations}\label{RelationsTotalComponents}
\begin{equation}\label{RelationsTotalComponents1}
\Arrowvert\tilde{\Phi}_{\tilde{\bm{u}}}\Arrowvert = \sqrt{\Arrowvert\tilde{\Phi}_{\tilde{\bm{u}}_{\textup{V}}}\Arrowvert^2 + \Arrowvert\tilde{\Phi}_{\tilde{\bm{u}}_{\textup{T}}}\Arrowvert^2 + 2 \Arrowvert\tilde{\Phi}_{\tilde{\bm{u}}_{\textup{V}}}\Arrowvert \Arrowvert\tilde{\Phi}_{\tilde{\bm{u}}_{\textup{T}}}\Arrowvert \cos{(\beta)}} \, ,
\end{equation}
\begin{equation}\label{RelationsTotalComponents2}
\Arrowvert\tilde{\Gamma}_{\tilde{\bm{u}}}\Arrowvert = \sqrt{\Arrowvert\tilde{\Gamma}_{\tilde{\bm{u}}_{\textup{V}}}\Arrowvert^2 + \Arrowvert\tilde{\Gamma}_{\tilde{\bm{u}}_{\textup{N}}}\Arrowvert^2 + 2 \Arrowvert\tilde{\Gamma}_{\tilde{\bm{u}}_{\textup{V}}}\Arrowvert \Arrowvert\tilde{\Gamma}_{\tilde{\bm{u}}_{\textup{N}}}\Arrowvert \sin{(\beta)}} \, .
\end{equation}
\end{subequations}
The ratio of the length of the minor semi-axis to the length of the major semi-axis of the ellipse can also be expressed as a function of the flux and circulation of the displacement components:
\begin{equation}\label{RelationsChiComponents}
\chi = \frac{\Arrowvert\tilde{\Phi}_{\tilde{\bm{u}}_{\textup{V}}}\Arrowvert}{\Arrowvert\tilde{\Phi}_{\tilde{\bm{u}}_{\textup{V}}}\Arrowvert + \sqrt{\Arrowvert\tilde{\Phi}_{\tilde{\bm{u}}_{\textup{T}}}\Arrowvert^2 + \Arrowvert\tilde{\Gamma}_{\tilde{\bm{u}}_{\textup{N}}}\Arrowvert^2}} \, .
\end{equation}

\subsection{Wave propagation at the stationary points of the dispersion surfaces}
\label{StationaryPointsChiralSection}

\begin{figure}%[!h]
\centering
\includegraphics[width=0.7\columnwidth]{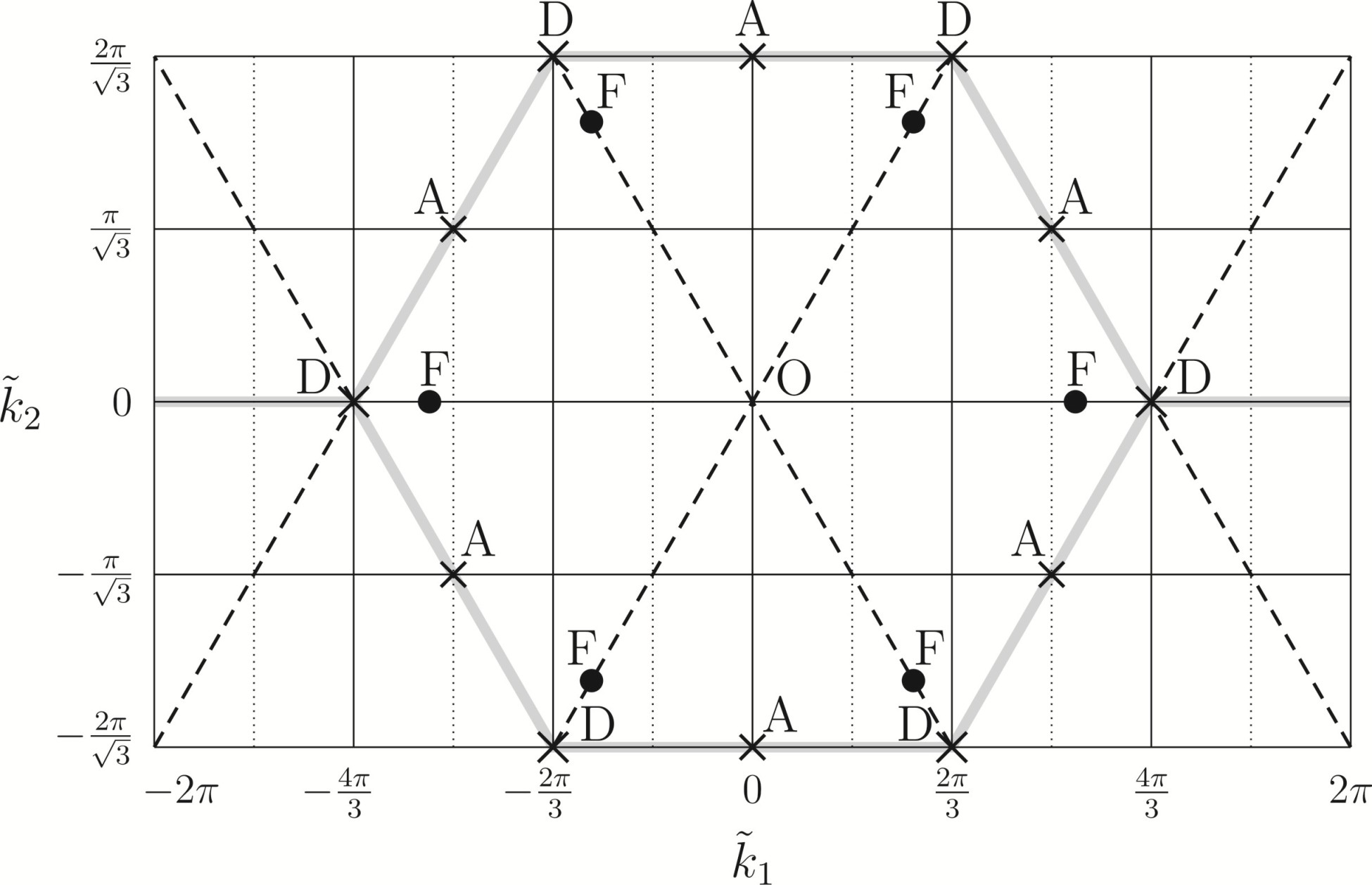}
\caption{\footnotesize Stationary points of the lower dispersion surface (A, D) and upper dispersion surface (A, D, F), whose properties are detailed in Tables \ref{Table1}-\ref{Table2c}, are shown in the extended rectangular domain which includes the hexagonal cell in the reciprocal space. The crosses represent fixed points, while the dots indicate points whose positions vary with $\tilde{\alpha}$. In the figure, the positions of points F are given for $\tilde{\alpha}=0$.}
\label{StationaryPoints}
\end{figure}

The stationary points of the two dispersion surfaces of the chiral lattice are shown in Fig. \ref{StationaryPoints}. The properties of the stationary points of the lower dispersion surface are detailed in Table \ref{Table1} for any value of the spinner constant $\tilde{\alpha}$, while those of the upper dispersion surface are given in Tables \ref{Table2a}, \ref{Table2b} and \ref{Table2c} for different ranges of $\tilde{\alpha}$. The coordinates of only one instance of each type of stationary point are detailed in Tables \ref{Table1}-\ref{Table2c}; the coordinates of the other corresponding stationary points can be obtained by rotating the given coordinates by $n \pi/3$ ($n=1,2,..,5$) with respect to the origin O.

It is interesting to note that points A and D do not change their positions in the $\tilde{\bm{k}}$-plane as $\tilde{\alpha}$ is changed, while positions of points F (which are stationary points only for the upper surface) are $\tilde{\alpha}$ dependent. In particular, for $\tilde{\alpha}=0$, points F occupy the positions shown in Fig. \ref{StationaryPoints}. When $\tilde{\alpha} = 1/3$, they coincide with points D; hence, for the upper dispersion surface, points D are minima for $\tilde{\alpha} < 1/3$, saddle points for $\tilde{\alpha} = 1/3$ and maxima for $1/3 < \tilde{\alpha} < 1$. When $\tilde{\alpha} \ge \sqrt{7/27}$, points F coincide with points A; accordingly, for the upper dispersion surface, points A are maxima for $\tilde{\alpha} < \sqrt{7/27}$ and become saddle points for $\sqrt{7/27} \le \tilde{\alpha} < 1$. Therefore, $\tilde{\alpha} = 1/3$ and $\tilde{\alpha} = \sqrt{7/27}$ are special values of the spinner constant, for which the response of the lattice changes significantly in terms of dynamic anisotropy.

The frequency of each stationary point varies with the spinner constant $\tilde{\alpha}$. The type of stationary point on the upper dispersion surface is also dependent on the spinner constant.

While in the non-chiral case ($\tilde{\alpha} = 0$) points F were on special lines characterised by either zero flux or zero circulation (see Fig. 7b in \cite{Carta2018}), in the chiral case both the flux and circulation at points F are generally different from zero. Conversely, points A and D are characterised by zero flux and zero circulation, since $\tilde{\bm{f}} = {\bf{0}}$ at these points. Additionally,  $\Arrowvert\tilde{\Phi}_{\tilde{\bm{u}}}^{(1)}\Arrowvert/\Arrowvert\tilde{\Gamma}_{\tilde{\bm{u}}}^{(1)}\Arrowvert < 1 < \Arrowvert\tilde{\Phi}_{\tilde{\bm{u}}}^{(2)}\Arrowvert/\Arrowvert\tilde{\Gamma}_{\tilde{\bm{u}}}^{(2)}\Arrowvert$ at points A, while  $\Arrowvert\tilde{\Phi}_{\tilde{\bm{u}}}^{(1)}\Arrowvert/\Arrowvert\tilde{\Gamma}_{\tilde{\bm{u}}}^{(1)}\Arrowvert,  \Arrowvert\tilde{\Phi}_{\tilde{\bm{u}}}^{(2)}\Arrowvert/\Arrowvert\tilde{\Gamma}_{\tilde{\bm{u}}}^{(2)}\Arrowvert \to 1$ at points D. Therefore, $\chi^{(j)}=1$ ($j=1,2$) at points D (see also Figs. \ref{ChijChiral}a and \ref{ChijChiral}b) and hence the corresponding motion of each lattice particle is circular.

%%%%%%%%%%%%%%%%%%%%%%%%%%%%%%%%%%%%%%%%%%%%
%%%%%%%%%%%%%%%%%%%%%%%%%%%%%%%%%%%%%%%%%%%%

\renewcommand{\arraystretch}{3}
\begin{table}
\begin{center}
\caption{Stationary points of the lower dispersion surface for $0 < \tilde{\alpha} < \infty$.}
\label{Table1}
\begin{tabular}{c c c c}
\hline
Point & $\left(\tilde{k}_1, \tilde{k}_2\right)$ & $\tilde{\omega}_1$ & Type \\
\hline
A & $\left(\pi, \frac{\pi}{\sqrt{3}}\right)$ & $\sqrt{\frac{6}{2+\sqrt{1+3\tilde{\alpha}^2}}}$ & saddle point \\
D & $\left(\frac{4\pi}{3}, 0\right)$ & $\sqrt{\frac{9}{2(1+\tilde{\alpha})}}$ & maximum \\
\hline
\end{tabular}
\end{center}
\end{table}

\begin{table}
\begin{center}
\caption{Stationary points of the upper dispersion surface for $0 < \tilde{\alpha} < 1/3$.}
\label{Table2a}
\begin{tabular}{c c c c}
\hline
Point & $\left(\tilde{k}_1, \tilde{k}_2\right)$ & $\tilde{\omega}_2$ & Type  \\
\hline
A & $\left(\pi, \frac{\pi}{\sqrt{3}}\right)$ & $\sqrt{\frac{6}{2-\sqrt{1+3\tilde{\alpha}^2}}}$ & maximum \\
D & $\left(\frac{4\pi}{3}, 0\right)$ & $\sqrt{\frac{9}{2(1-\tilde{\alpha})}}$ & minimum \\
F & $\left(4\arccos{\left[\frac{\sqrt{7-27\tilde{\alpha}^2}}{4}\right]}, 0\right)$ & $\frac{9\sqrt{1+3\tilde{\alpha}^2}}{4}$ & saddle point \\
\hline
\end{tabular}
\end{center}
\end{table}

\begin{table}
\begin{center}
\caption{Stationary points of the upper dispersion surface for $1/3 < \tilde{\alpha} < \sqrt{7/27}$.}
\label{Table2b}
\begin{tabular}{c c c c}
\hline
Point & $\left(\tilde{k}_1, \tilde{k}_2\right)$ & $\tilde{\omega}_2$ & Type  \\
\hline
A & $\left(\pi, \frac{\pi}{\sqrt{3}}\right)$ & $\sqrt{\frac{6}{2-\sqrt{1+3\tilde{\alpha}^2}}}$ & maximum \\
D & $\left(\frac{4\pi}{3}, 0\right)$ & $\sqrt{\frac{9}{2(1-\tilde{\alpha})}}$ & maximum \\
F & $\left(2 \pi - \arccos{\left[\frac{-1-27 \tilde{\alpha}^2}{8}\right]}, \sqrt{3}\arccos{\left[\frac{-1-27 \tilde{\alpha}^2}{8}\right]}-\frac{2 \pi}{\sqrt{3}}\right)$ & $\frac{9\sqrt{1+3\tilde{\alpha}^2}}{4}$ & saddle point \\
\hline
\end{tabular}
\end{center}
\end{table}

\begin{table}
\begin{center}
\caption{Stationary points of the upper dispersion surface for $\sqrt{7/27} < \tilde{\alpha} < 1$.}
\label{Table2c}
\begin{tabular}{c c c c}
\hline
Point & $\left(\tilde{k}_1, \tilde{k}_2\right)$ & $\tilde{\omega}_2$ & Type  \\
\hline
A & $\left(\pi, \frac{\pi}{\sqrt{3}}\right)$ & $\sqrt{\frac{6}{2-\sqrt{1+3\tilde{\alpha}^2}}}$ & saddle point \\
D & $\left(\frac{4\pi}{3}, 0\right)$ & $\sqrt{\frac{9}{2(1-\tilde{\alpha})}}$ & maximum \\
F & $\left(\pi, \frac{\pi}{\sqrt{3}}\right)$ & $\frac{9\sqrt{1+3\tilde{\alpha}^2}}{4}$ & saddle point \\
\hline
\end{tabular}
\end{center}
\end{table}

\section{Illustrative examples and physical interpretation of wave characterisation}
\label{ExamplesChiralSection}

In this section, we investigate how waves propagate in the chiral medium for different values of the wave vector. In particular, we show the total displacement field of the lattice in time, as well as the motion of the lattice particles when a single component of the displacement field (vortex, straight-line, straight-line tangential or straight-line normal) is considered. In the calculations, the spinner constant is taken as $\tilde{\alpha} = 0.5$. We emphasise that increasing $\tilde{\alpha}$ augments the contribution of the vortex component to the total field and makes the elliptical trajectories of the lattice particles less eccentric.

\begin{figure}%[!h]
\centering
\includegraphics[width=1.0\columnwidth]{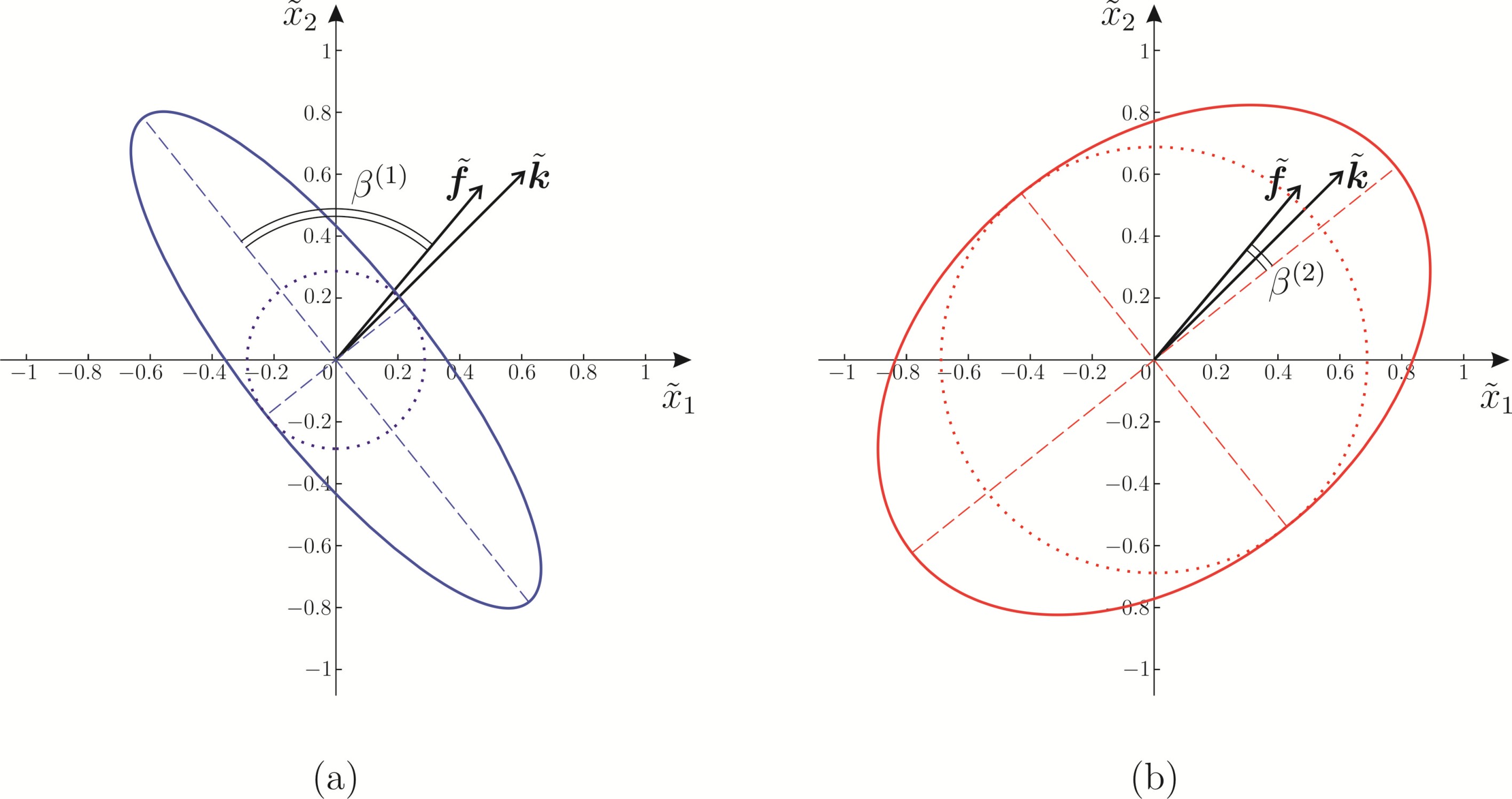}
\caption{\footnotesize Trajectory of a generic particle in the chiral lattice with spinner constant $\tilde{\alpha}=0.5$ corresponding to the (a) lower and (b) upper dispersion surface, calculated for $\tilde{k}_1 = 2$ and $\tilde{k}_2 = 2$. The dotted circles correspond to the vortex components of the displacements.}
\label{FigureExampleChiral}
\end{figure}

Firstly, we consider a relatively large value of $\left|\tilde{\bm{k}}\right|$, namely $\tilde{k}_1 = 2$ and $\tilde{k}_2 = 2$. The vector $\tilde{\bm{f}}$, defined in (\ref{vectorf}), is given by $\tilde{\bm{f}} = \left( 1.548,1.847 \right)^{\mathrm{T}}$. The corresponding frequencies for the lower and upper dispersion surface are $\tilde{\omega}^{(1)} = 1.360$ and $\tilde{\omega}^{(2)} = 2.779$, respectively. The angles between the vector $\tilde{\bm{f}}$ and the major axes of the elliptical trajectories of the lattice particles for the two dispersion surfaces are $\beta^{(1)} = 1.370$ and $\beta^{(2)} = 0.200$. We remark that, because of the choice of decomposition of the displacement, the angles $\beta^{(j)}$ ($j=1,2$) do not vary with $\tilde{\alpha}$; changing the value of $\tilde{\alpha}$ changes only the vortex component. The amplitudes of flux and circulation corresponding to the lower surface are given by $\Arrowvert\tilde{\Phi}_{\tilde{\bm{u}}}^{(1)}\Arrowvert = 0.720$ and $\Arrowvert\tilde{\Gamma}_{\tilde{\bm{u}}}^{(1)}\Arrowvert = 2.049$, while those obtained for the upper surface are $\Arrowvert\tilde{\Phi}_{\tilde{\bm{u}}}^{(2)}\Arrowvert = 2.066$ and $\Arrowvert\tilde{\Gamma}_{\tilde{\bm{u}}}^{(2)}\Arrowvert = 1.468$. The ratios of flux to circulation are respectively $\Arrowvert\tilde{\Phi}_{\tilde{\bm{u}}}^{(1)}\Arrowvert / \Arrowvert\tilde{\Gamma}_{\tilde{\bm{u}}}^{(1)}\Arrowvert = 0.351$ and $\Arrowvert\tilde{\Phi}_{\tilde{\bm{u}}}^{(2)}\Arrowvert / \Arrowvert\tilde{\Gamma}_{\tilde{\bm{u}}}^{(2)}\Arrowvert = 1.407$. The values above show that waves in this chiral system are a mixture of flux-free, circulation-free and vortex contributions. The ratios between the lengths of the minor semi-axes to the lengths of the major semi-axes of the ellipse for the lower and upper surface are $\chi^{(1)} = 0.287$ and $\chi^{(2)} = 0.688$, respectively. These values of $\chi^{(j)}$ are associated with the amplitudes of flux and circulation of the vortex motion, which are $\Arrowvert\tilde{\Phi}_{\tilde{\bm{u}}_{\textup{V}}}^{(1)}\Arrowvert=\Arrowvert\tilde{\Gamma}_{\tilde{\bm{u}}_{\textup{V}}}^{(1)}\Arrowvert=0.600$ for the lower dispersion surface and $\Arrowvert\tilde{\Phi}_{\tilde{\bm{u}}_{\textup{V}}}^{(2)}\Arrowvert=\Arrowvert\tilde{\Gamma}_{\tilde{\bm{u}}_{\textup{V}}}^{(2)}\Arrowvert=1.437$ for the upper dispersion surface. The flux and circulation corresponding to the straight-line component are $\Arrowvert\tilde{\Phi}_{\tilde{\bm{u}}_{\textup{T}}}^{(1)}\Arrowvert=0.296$, $\Arrowvert\tilde{\Gamma}_{\tilde{\bm{u}}_{\textup{N}}}^{(1)}\Arrowvert=1.458$ for the lower surface and $\Arrowvert\tilde{\Phi}_{\tilde{\bm{u}}_{\textup{T}}}^{(2)}\Arrowvert=0.637$, $\Arrowvert\tilde{\Gamma}_{\tilde{\bm{u}}_{\textup{N}}}^{(2)}\Arrowvert=0.129$ for the upper surface. The elliptical trajectory of each lattice particle is illustrated in Fig. \ref{FigureExampleChiral}a and Fig. \ref{FigureExampleChiral}b for the lower and upper dispersion surface, respectively. The circles in Fig. \ref{FigureExampleChiral} correspond to the vortex components of the displacement (compare with Fig. \ref{EllipticalTrajectoryChiral}).

%%%%%%%%%%%%%%%%%%%%%%%%%%%%%%%%%%%%%%%%%%%%
%%%%%%%%%%%%%%%%%%%%%%%%%%%%%%%%%%%%%%%%%%%%

%Video 1a and Video 1f in the Supplementary Material show the total time-dependent displacement fields corresponding to the two dispersion surfaces, determined for the wave vector with components $\tilde{k}_1 = 2$ and $\tilde{k}_2 = 2$. The contributions due to the vortex components are illustrated in Video 1b and Video 1g, while those due to the straight-line components are displayed in Video 1c and Video 1h. The straight-line field is further decomposed into a flux-free (normal) motion (see Video 1d and Video 1i) and a circulation-free (tangential) motion (see Video 1e and Video 1j). In the videos, the amplitude of each displacement component is proportional to its contribution to the total displacement. Accordingly, it can be observed that for the upper surface the vortex component plays a significant role for this choice of parameters. In the videos, the vectors $\tilde{\bm{k}}$ and $\tilde{\bm{f}}$ are indicated in magenta and green, respectively, and the trajectories of the particles are plotted in red.

\begin{figure}%[!h]
\centering
\includegraphics[width=0.9\columnwidth]{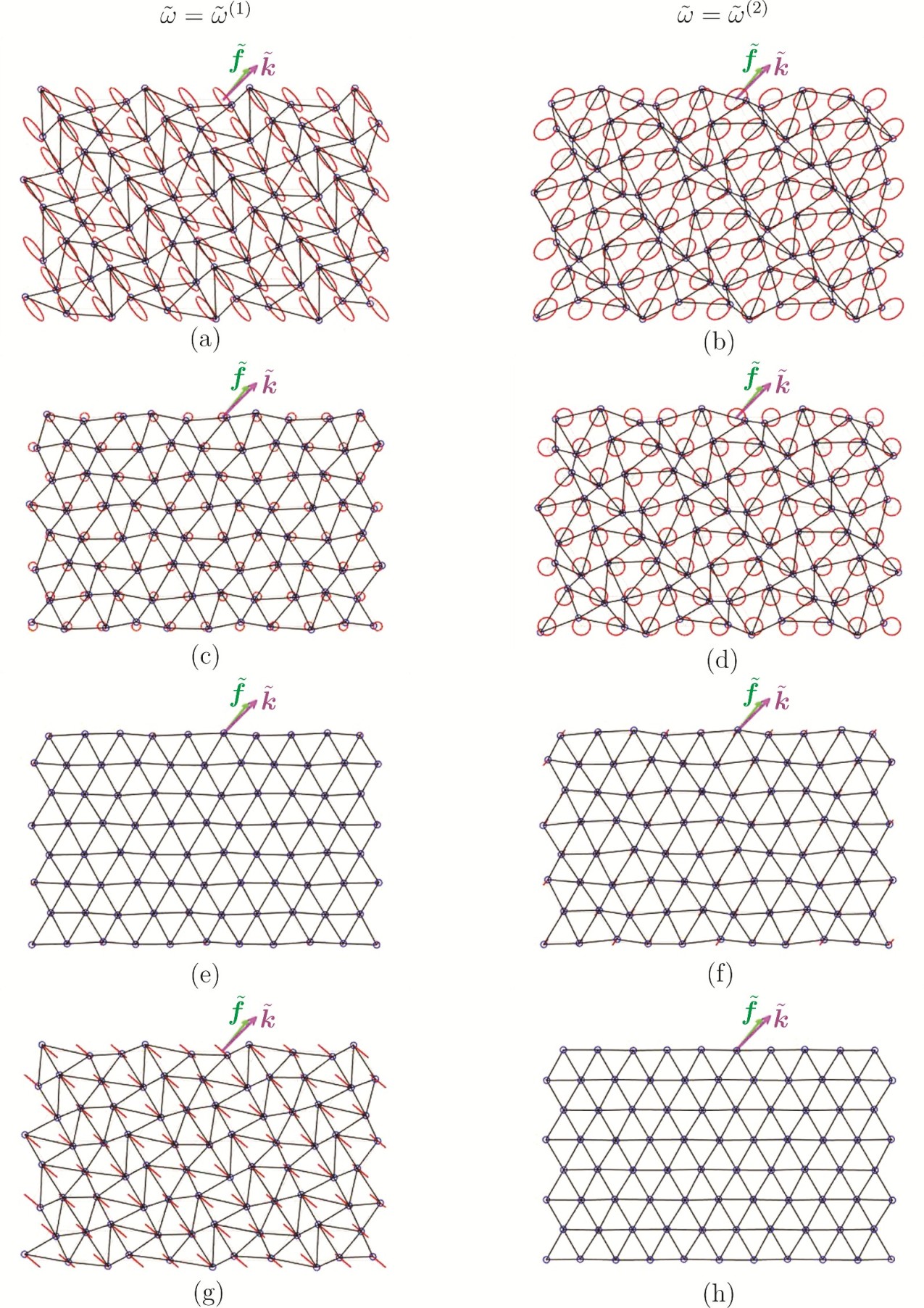}
\caption{\footnotesize Displacement field in the chiral lattice for the (a,c,e,g) lower and (b,d,f,h) upper dispersion surface, calculated for $\tilde{k}_1 = 2$, $\tilde{k}_2 = 2$ and $\tilde{\alpha} = 0.5$. (a,b) Total displacements, (c,d) vortex components, (e,f) straight-line tangential components, (g,h) straight-line normal components. The vectors $\tilde{\bm{k}}$ and $\tilde{\bm{f}}$ are also plotted. The trajectories of the lattice particles are shown in red.}
\label{Snapshots1}
\end{figure}

Figs. \ref{Snapshots1}a and \ref{Snapshots1}b show instantaneous snapshots of the displacement fields for $\tilde{k}_1 = 2$, $\tilde{k}_2 = 2$ calculated for the lower and upper dispersion surface, respectively. The corresponding vortex components are presented in Figs. \ref{Snapshots1}c and \ref{Snapshots1}d, the straight-line tangential components in Figs. \ref{Snapshots1}e and \ref{Snapshots1}f and the straight-line normal components in Figs. \ref{Snapshots1}g and \ref{Snapshots1}h.

%%%%%%%%%%%%%%%%%%%%%%%%%%%%%%%%%%%%%%%%%%%%
\begin{figure}%[!h]
\centering
\includegraphics[width=0.9\columnwidth]{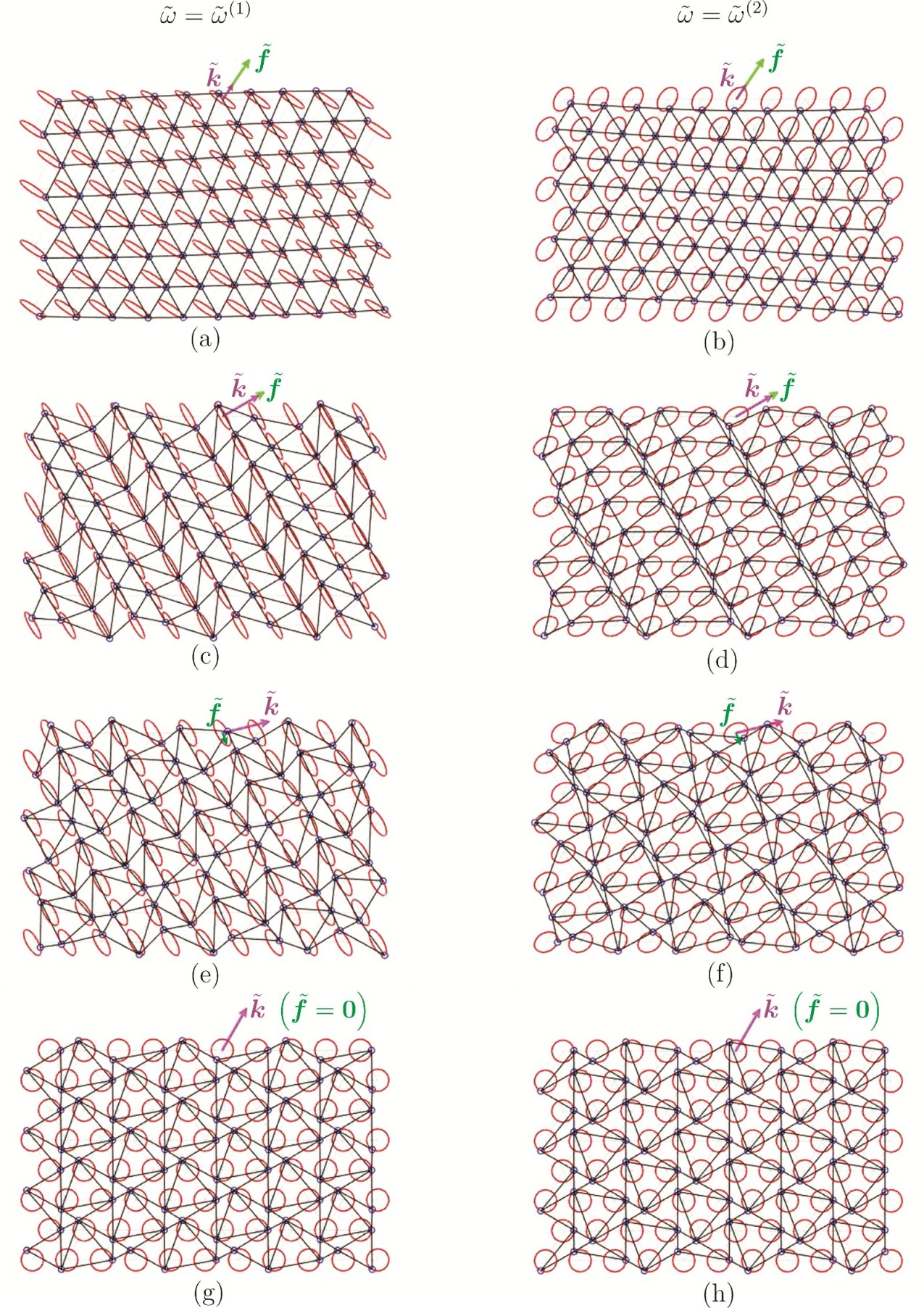}
\caption{\footnotesize Displacements fields in the chiral lattice with spinner constant $\tilde{\alpha} = 0.5$ relative to the (a,c,e,g) lower and (b,d,f,h) upper dispersion surface, calculated for different values of the wave vector: (a,b) $\tilde{k}_1 = 0.200$, $\tilde{k}_2 = 0.297$; (c,d) $\tilde{k}_1 = 2.150$, $\tilde{k}_2 = 1.241$; (e,f) $\tilde{k}_1 = 3.665$, $\tilde{k}_2 = 0.907$; (g,h) $\tilde{k}_1 = 2\pi/3$, $\tilde{k}_2 = 2\pi/\sqrt{3}$.}
\label{Snapshots2345}
\end{figure}
%%%%%%%%%%%%%%%%%%%%%%%%%%%%%%%%%%%%%%%%%%%%

We note that the rotations of the lattice particles corresponding to the two dispersion surfaces are always in opposite directions.

\section{Dynamic degeneracy in chiral elastic systems}
\label{SectionLimitValuesAlpha}

In this section, we study the factor $\chi^{(j)}$ for the lower ($j=1$) and upper ($j=2$) dispersion surface for some limit cases of the spinner constant $\tilde{\alpha}$. In addition, we discuss the possibility of creating vortex waveforms for any value of the wave vector when $\tilde{\alpha}$ tends to unity.

\subsection{Lower dispersion surface}
\label{SectionLimitValuesAlphaLower}

\begin{figure}%[!h]
\centering
\includegraphics[width=0.65\columnwidth]{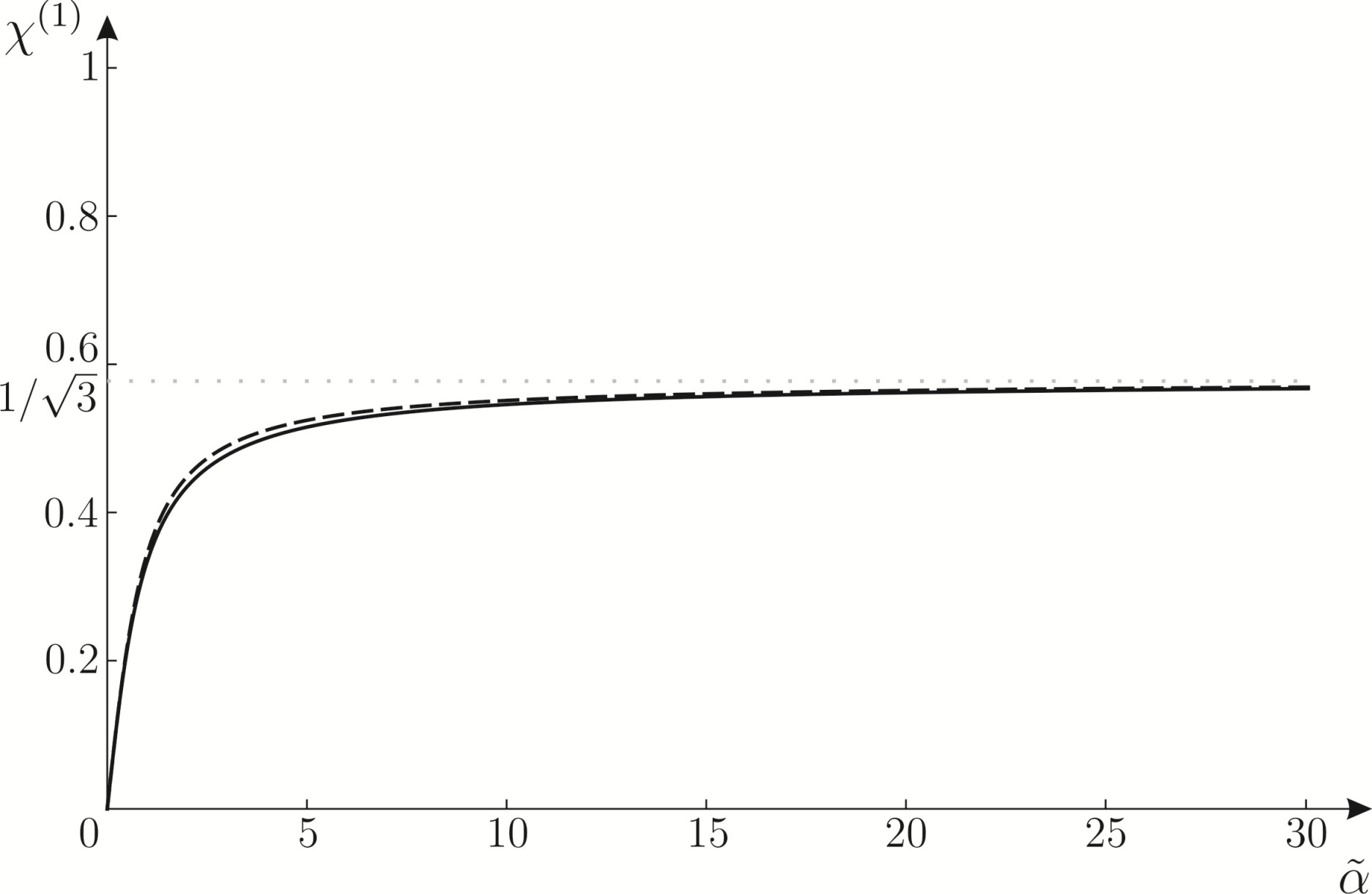}
\caption{\footnotesize For the long wavelength limit, graph of $\chi^{(1)}$ versus $\tilde{\alpha}$ (solid line) together with its approximation (\ref{Chi1AlphaArcTan}) (dashed line).}
\label{Chi1Alpha}
\end{figure}

Consider the long wavelength limit, when $\left|\tilde{\bm{k}}\right| \to 0$. In this limit, Eq. (\ref{RelationsChiComponents}) leads to
\begin{equation}\label{chi1smallk}
\chi^{(1)} \sim \sqrt{\frac{2 + \frac{1+\sqrt{1+3\tilde{\alpha}^2}}{\tilde{\alpha}^2} - \sqrt{\frac{\tilde{\alpha}^4+2\left(1+\sqrt{1+3\tilde{\alpha}^2}\right)+\tilde{\alpha}^2\left(5+2\sqrt{1+3\tilde{\alpha}^2}\right)}{\tilde{\alpha}^4}}}{2 + \frac{1+\sqrt{1+3\tilde{\alpha}^2}}{\tilde{\alpha}^2} + \sqrt{\frac{\tilde{\alpha}^4+2\left(1+\sqrt{1+3\tilde{\alpha}^2}\right)+\tilde{\alpha}^2\left(5+2\sqrt{1+3\tilde{\alpha}^2}\right)}{\tilde{\alpha}^4}}}} \quad \text{when } \left|\tilde{\bm{k}}\right| \to 0 \, .
\end{equation}
The function above is shown in Fig. \ref{Chi1Alpha} by a solid line.

%%%%%%%%%%%%%%%%%%%%%%%%%%%%%%%%%%%%%%%%%%%%
%%%%%%%%%%%%%%%%%%%%%%%%%%%%%%%%%%%%%%%%%%%%

Since $\chi^{(1)} \sim \tilde{\alpha}/2$ when $\tilde{\alpha} \to 0$ and $\chi^{(1)} \to 1/\sqrt{3}$ when $\tilde{\alpha} \to \infty$, the expression (\ref{chi1smallk}) can be approximated by the simpler function
\begin{equation}\label{Chi1AlphaArcTan}
\chi^{(1)} \approx \frac{2}{\sqrt{3}\pi} \arctan{\left( \frac{\sqrt{3}\pi}{4} \tilde{\alpha} \right)} \, ,
\end{equation}
which is represented by a dashed line in Fig. \ref{Chi1Alpha}.

The limit $\chi^{(1)} \to 1/\sqrt{3}$ when $\tilde{\alpha} \to \infty$ can be proved as follows. For simplicity and without loss of generality, we take again $\mathrm{Re}( \tilde{U}_2^{(1)} ) = 1$ and $\mathrm{Im}( \tilde{U}_2^{(1)} ) = 0$ in (\ref{Displacements}). The frequency $\tilde{\omega}^{(1)}$, corresponding to the lower dispersion surface and given by (\ref{Omega12}a), has the following asymptotic expansion for large values of the spinner constant and in the long wavelength limit:
\begin{equation}\label{omega1asympt}
\tilde{\omega}^{(1)} \sim \frac{\sqrt[4]{27}}{2\sqrt{2}} \left( \frac{\tilde{k}_1^2}{2 \tilde{k}_2} + \tilde{k}_2 \right) \sqrt{\epsilon} \quad \text{when } \left|\tilde{\bm{k}}\right| \to 0 \;\; \text{and } \epsilon = \frac{1}{\tilde{\alpha}} \to 0 \, .
\end{equation}
Substituting the above expression into Eqs. (\ref{EquationsOfMotion}) to determine the eigenvectors, we find that $\mathrm{Re}( \tilde{U}_1^{(1)} )$ and $\mathrm{Im}( \tilde{U}_1^{(1)} )$ in (\ref{Displacements}) are
\begin{equation}\label{Ur1Ui1asympt}
\begin{split}
\mathrm{Re}( \tilde{U}_1^{(1)} ) &\sim -\frac{2\tilde{k}_1\tilde{k}_2}{3\tilde{k}_1^2+\tilde{k}_2^2} \, , \; \mathrm{Im}( \tilde{U}_1^{(1)} ) \sim \frac{\sqrt{3}\left(\tilde{k}_1^2+\tilde{k}_2^2\right)}{3\tilde{k}_1^2+\tilde{k}_2^2} \\
&\text{when } \left|\tilde{\bm{k}}\right| \to 0 \;\; \text{and } \epsilon = \frac{1}{\tilde{\alpha}} \to 0 \, .
\end{split}
\end{equation}
The eigenvalues (\ref{EigenvaluesEllipse}) are found to be
\begin{equation}\label{lambdasasympt}
\lambda_{-}^{(1)} \sim \frac{3\tilde{k}_1^2+\tilde{k}_2^2}{3\left(\tilde{k}_1^2+\tilde{k}_2^2\right)} \, , \; \lambda_{+}^{(1)} \sim \frac{3\tilde{k}_1^2+\tilde{k}_2^2}{\tilde{k}_1^2+\tilde{k}_2^2} \quad \text{when } \left|\tilde{\bm{k}}\right| \to 0 \;\; \text{and } \epsilon = \frac{1}{\tilde{\alpha}} \to 0 \, .
\end{equation}
Hence, Eq. (\ref{ChiralityFactor}) leads to
\begin{equation}\label{chi1asympt}
\chi^{(1)} = \sqrt{\frac{\lambda_{-}^{(1)}}{\lambda_{+}^{(1)}}} \sim \frac{1}{\sqrt{3}} \quad \text{when } \left|\tilde{\bm{k}}\right| \to 0 \;\; \text{and } \epsilon = \frac{1}{\tilde{\alpha}} \to 0 \, .
\end{equation}
The same limit for $\tilde{\alpha} \to \infty$ is attained by the approximation (\ref{Chi1AlphaArcTan}).

For large values of the wave vector, in the limit when $\tilde{\alpha} \to \infty$, $\chi^{(1)}$ is given by
\begin{equation}\label{chi2asympt2}
\chi^{(1)} = \chi^{(1)} \left( \tilde{k}_1, \tilde{k}_2 \right) \sim \sqrt{\frac{1 + \cal{N}_1/\cal{D}^2 - \cal{N}_2/\cal{D}}{1 + \cal{N}_1/\cal{D}^2 + \cal{N}_2/\cal{D}}} \quad \text{when } \tilde{\alpha} \to \infty \, ,
\end{equation}
where
\begin{subequations}\label{chi2asympt2N1N2D}
\begin{align}
\cal{N}_1 & = 12 \left( 5 - 4 c_1^2 - 6 c_1 c_2 + c_1^2 c_2^2 + 4 c_1^3 c_2 \right) \, , \\
\cal{N}_2 & = 4 \sqrt{4 - 7 c_1^2 + 4 c_1^4 + 2 c_1 c_2 - 3 c_2^2 + 4 c_1^2 c_2^2 - 4 c_1^3 c_2} \, , \\
\cal{D} & = 10 - 8 c_1^2 - 2 c_1 c_2 \, ,
\end{align}
\end{subequations}
with $c_1 = \cos{\left(\frac{\tilde{k}_1}{2}\right)}$ and $c_2 = \cos{\left(\frac{\sqrt{3}\tilde{k}_2}{2}\right)}$. We note that, in the unit cell, $\cal{D}=0$ for $\left( \tilde{k}_1,\tilde{k}_2 \right)^{\textup{T}} = \left( 0,0 \right)^{\textup{T}}$ and $\left( \tilde{k}_1,\tilde{k}_2 \right)^{\textup{T}} = \left( \pm 2\pi,\pm 2\pi/\sqrt{3} \right)^{\textup{T}}$. Nonetheless, $\cal{N}_1/\cal{D}^2 \to 1$ and $\cal{N}_2/\cal{D} \to 1$ for $\left( \tilde{k}_1,\tilde{k}_2 \right)^{\textup{T}} \to \left( 0,0 \right)^{\textup{T}}$ and $\left( \tilde{k}_1,\tilde{k}_2 \right)^{\textup{T}} \to \left( \pm 2\pi,\pm 2\pi/\sqrt{3} \right)^{\textup{T}}$.

\begin{figure}%[!h]
\centering
\includegraphics[width=0.7\columnwidth]{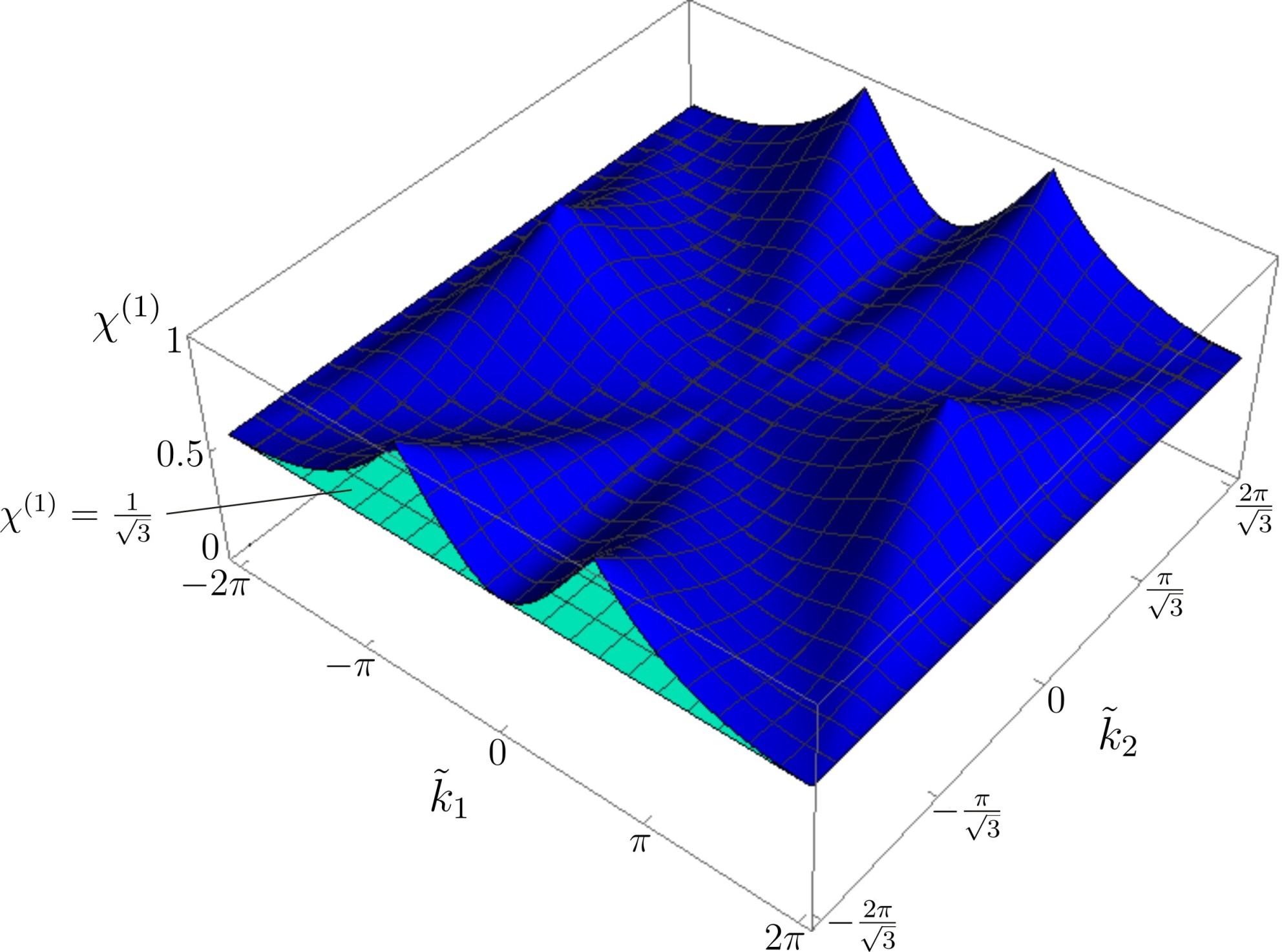}
\caption{\footnotesize $\chi^{(1)}$ in the $\tilde{\bm{k}}$-plane in the limit when $\tilde{\alpha} \to \infty$. The plane $\chi^{(1)} = 1/\sqrt{3}$ is also included in the figure.}
\label{Chi13D}
\end{figure}

The function $\chi^{(1)} \left( \tilde{k}_1, \tilde{k}_2 \right)$ given in (\ref{chi2asympt2}) is shown in Fig. \ref{Chi13D}. We observe that the global minimum of the function is $1/\sqrt{3}$, that is the value obtained for $\left|\tilde{\bm{k}}\right| \to 0$ (see Eq. (\ref{chi1asympt})). Furthermore, $\chi^{(1)} = 1/\sqrt{3}$ along the radials given by $\arctan{\left( \tilde{k}_2/\tilde{k}_1 \right)} = (2n-1)\pi/6$ with $n=1,..,6$. The global maxima are found at the points D in Fig. \ref{StationaryPoints}, where $\chi^{(1)} = 1$.

%%%%%%%%%%%%%%%%%%%%%%%%%%%%%%%%%%%%%%%%%%%%
%%%%%%%%%%%%%%%%%%%%%%%%%%%%%%%%%%%%%%%%%%%%

\subsection{Upper dispersion surface}
\label{SectionLimitValuesAlphaUpper}

\begin{figure}%[!h]
\centering
\includegraphics[width=0.65\columnwidth]{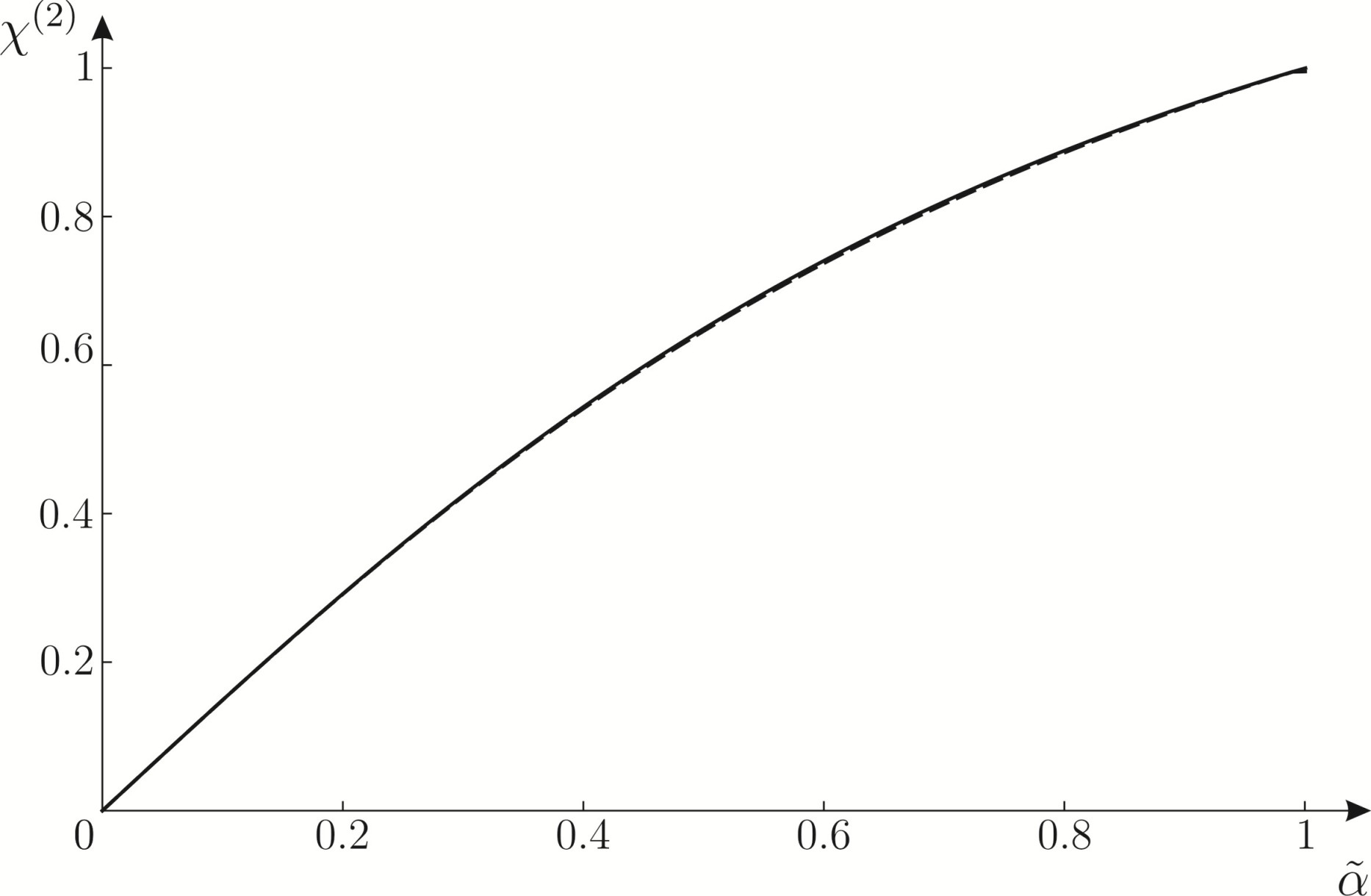}
\caption{\footnotesize For $\left|\tilde{\bm{k}}\right| \to 0$, $\chi^{(2)}$ versus $\tilde{\alpha}$ (solid line), compared with the analytical approximation (\ref{Chi2AlphaArcTan}) (dashed line).}
\label{Chi2Alpha}
\end{figure}

As in Section \ref{SectionLimitValuesAlpha}\ref{SectionLimitValuesAlphaLower}, we first analyse the case when $\left|\tilde{\bm{k}}\right| \to 0$. Using (\ref{RelationsChiComponents}), we find that
\begin{equation}\label{chi2smallk}
\chi^{(1)} \sim \sqrt{\frac{6 + \frac{1+\sqrt{1+3\tilde{\alpha}^2}}{\tilde{\alpha}^2} - \sqrt{\frac{9\tilde{\alpha}^4+2\left(1+\sqrt{1+3\tilde{\alpha}^2}\right)-3\tilde{\alpha}^2\left(1+2\sqrt{1+3\tilde{\alpha}^2}\right)}{\tilde{\alpha}^4}}}{6 + \frac{1+\sqrt{1+3\tilde{\alpha}^2}}{\tilde{\alpha}^2} + \sqrt{\frac{9\tilde{\alpha}^4+2\left(1+\sqrt{1+3\tilde{\alpha}^2}\right)-3\tilde{\alpha}^2\left(1+2\sqrt{1+3\tilde{\alpha}^2}\right)}{\tilde{\alpha}^4}}}} \quad \text{when } \left|\tilde{\bm{k}}\right| \to 0 \, .
\end{equation}
The function above is shown in Fig. \ref{Chi2Alpha} by a solid line. In this case $0 \le \tilde{\alpha} < 1$, since $\tilde{\omega}^{(2)}$ takes imaginary values for $\tilde{\alpha} > 1$ (see Eq. (\ref{Omega12}b)).

%%%%%%%%%%%%%%%%%%%%%%%%%%%%%%%%%%%%%%%%%%%%
%%%%%%%%%%%%%%%%%%%%%%%%%%%%%%%%%%%%%%%%%%%%

In Fig. \ref{Chi2Alpha}, the dashed line represents the following analytical approximation of $\chi^{(2)}$:
\begin{equation}\label{Chi2AlphaArcTan}
\chi^{(2)} \approx \cal{C} \arctan{\left( \frac{3}{2 \cal{C}} \tilde{\alpha} \right)} \, ,
\end{equation}
where $\cal{C}$ is the root of the equation
\begin{equation}\label{Chi2AlphaArcTanC}
\cal{C} \arctan{\left( \frac{3}{2 \cal{C}} \right)} = 1 \, ,
\end{equation}
which gives $\cal{C} \approx 1.0337$. The approximation (\ref{Chi2AlphaArcTan}) has the same limits as the function (\ref{chi2smallk}) for small and large values of $\tilde{\alpha}$, namely $3\tilde{\alpha}/2$ when $\tilde{\alpha} \to 0$ and $1$ when $\tilde{\alpha} \to 1$.

We note that, as discussed below, for any $\tilde{\bm{k}}$ (not only in the long wavelength limit) we have
\begin{equation}\label{limit}
\lim_{\tilde{\alpha} \to 1} \chi^{(2)} = 1 \, .
\end{equation}

The frequency on the upper dispersion surface has the following asymptotic approximation when $\tilde{\alpha} \to 1$:
\begin{equation}\label{omega2asympt}
\tilde{\omega}^{(2)} \sim \frac{\sqrt{3 - \cos{(\tilde{k}_1)} - 2 \cos{(\tilde{k}_1/2)}\cos{(\sqrt{3}\tilde{k}_2/2)}}}{\sqrt{\epsilon}} \quad \text{for } \tilde{\alpha} = 1-\epsilon \;\; \text{with } \epsilon \to 0^+ \, .
\end{equation}
Substituting the expression above into Eq. (\ref{EquationsOfMotion}) and normalising the eigenvector in (\ref{Displacements}) by setting $\mathrm{Re}( \tilde{U}_2^{(2)} ) = 1$ and $\mathrm{Im}( \tilde{U}_2^{(2)} ) = 0$, we obtain
\begin{equation}\label{Ur2Ui2asympt}
\mathrm{Re}( \tilde{U}_1^{(2)} ) \to 0 \, , \; \mathrm{Im}( \tilde{U}_1^{(2)} ) \to 1 \quad \text{for } \tilde{\alpha} = 1-\epsilon \;\; \text{with } \epsilon \to 0^+ \, .
\end{equation}
Consequently, from (\ref{EigenvaluesEllipse}) $\lambda_{-}^{(2)}$ and $\lambda_{+}^{(2)}$ are found to be equal, and hence from (\ref{ChiralityFactor}) we have
\begin{equation}\label{chi2asympt}
\chi^{(2)} = \sqrt{\frac{\lambda_{-}^{(2)}}{\lambda_{+}^{(2)}}} \to 1 \quad \text{for } \tilde{\alpha} = 1-\epsilon \;\; \text{with } \epsilon \to 0^+ \, .
\end{equation}

The results of this section demonstrate that pure vortex waveforms can be realised for any value of the wave vector at higher frequencies, when the spinner constant tends to its critical value ($\tilde{\alpha} \to 1$).

\section{Conclusions}
\label{Conclusions}

In this paper, we have demonstrated that the analytical concepts of lattice flux and lattice circulation represent canonical characteristics to describe polarisation of waves in a chiral elastic lattice. This is especially important when the wavelength is comparable with the size of the elementary cell of the periodic system, where the continuum notions of pressure and shear waves cannot be used.

The procedure discussed in this paper allows for a canonical decomposition of a general waveform in a chiral lattice. Besides flux-free and circulation-free straight-line displacement components, typical of the non-chiral case ($\tilde{\alpha} = 0$) discussed in \cite{Carta2018}, in a chiral lattice the concept of vortex waveforms has been introduced and investigated here.

Asymptotic analysis and animations have shown limit situations when vortex waveforms become dominant. In these cases, the trajectories of the lattice particles are circular and the amplitudes of lattice flux and lattice circulation are equal.

The analytical findings presented in this paper provide a method of interpreting waves in chiral elastic lattices.\vskip6pt

\vspace*{5mm}
\noindent
{\bf Acknowledgments }
{
\noindent
The authors would like to thank the EPSRC (UK) for its support through Programme Grant no. EP/L024926/1.
}


\begin{thebibliography}{50}

\bibitem{MartinssonMovchan2003}
Martinsson PG, Movchan AB. 2003. Vibrations of lattice structures and phononic band gaps. \textit{Q. J. Mech. Appl. Math.} \textbf{56}, 45--64. %(doi: 10.1093/qjmam/56.1.45)

\bibitem{Ayzenberg-StepanenkoSlepyan2008}
Ayzenberg-Stepanenko M, Slepyan LI. 2008. Resonant-frequency primitive waveforms and star waves in lattices. \textit{J. Sound Vib.} \textbf{313}, 812--821. %(doi: 10.1016/j.jsv.2007.11.047)

\bibitem{Colquitt2012}
Colquitt DJ, Jones IS, Movchan NV, Movchan AB, McPhedran RC. 2012. Dynamic anisotropy and localization in elastic lattice systems. \textit{Waves Random Complex Media} \textbf{22}, 143--159. %(doi: 10.1080/17455030.2011.633940)

\bibitem{MarderLiu1993}
Marder M, Liu X. 1993. Instability in lattice fracture. \textit{Phys. Rev. Lett.} \textbf{71}, 2417. %(doi: 10.1103/PhysRevLett.71.2417)

\bibitem{MarderGross1995}
Marder M, Gross S. 1995. Origin of crack tip instabilities. \textit{J. Mech. Phys. Solids} \textbf{43}, 1--48. %(doi: 10.1016/0022-5096(94)00060-I)

\bibitem{Slepyan2002}
Slepyan LI. 2002. \textit{Models and Phenomena in Fracture Mechanics}. Berlin: Springer.

\bibitem{Slepyan2010}
Slepyan LI, Movchan AB, Mishuris GS. 2010. Crack in a lattice waveguide. \textit{Int. J. Fract.} \textbf{162}, 91--106. %(doi: 10.1007/s10704-009-9389-5)

\bibitem{Nieves2016}
Nieves MJ, Mishuris GS, Slepyan LI. 2016. Analysis of dynamic damage propagation in discrete beam structures. \textit{Int. J. Solids Struct.} \textbf{97-98}, 699--713. %(doi: 10.1016/j.ijsolstr.2016.02.033)

\bibitem{Nieves2017}
Nieves MJ, Mishuris GS, Slepyan LI. 2017. Transient wave in a transformable periodic flexural structure. \textit{Int. J. Solids Struct.} \textbf{112}, 185--208. %(doi: 10.1016/j.ijsolstr.2016.11.012)

\bibitem{Bensoussan1978}
Bensoussan A, Lions JL, Papanicolaou G. 1978. \textit{Asymptotic Analysis for Periodic Structures}. Amsterdam: North-Holland.

\bibitem{BakhvalovPanasenko1984}
Bakhvalov NS, Panasenko GP. 1984. \textit{Homogenization: Averaging Processes in Periodic Media}. Mathematics and Its Applications (Soviet Series), vol. 36. Dordrecht-Boston-London: Kluwer Academic Publishers.

\bibitem{ZhikovKozlovOleinik1994}
Zhikov VV, Kozlov SM, Oleinik OA. 1994. \textit{Homogenization of Differential Operators and Integral Functionals}. Heidelberg: Springer.

\bibitem{Panasenko2005}
Panasenko GP. 2005. \textit{Multi-scale Modelling for Structures and Composites}. Dordrecht: Springer.

\bibitem{Craster2010}
Craster RV, Kaplunov J, Postnova J. 2010. High-frequency asymptotics, homogenisation and localisation for lattices. \textit{Q. J. Mech. Appl. Math.} \textbf{63}, 497--519. %(doi: 10.1093/qjmam/hbq015)

\bibitem{Antonakakis2014}
Antonakakis T, Craster RV, Guenneau S. 2014. Homogenisation for elastic photonic crystals and dynamic anisotropy. \textit{J. Mech. Phys. Solids} \textbf{71}, 84--96. %(doi: 10.1016/j.jmps.2014.06.006)

\bibitem{MovchanSlepyan2014}
Movchan AB, Slepyan LI. 2014. Resonant waves in elastic structured media: Dynamic homogenisation versus Green's functions. \textit{Int. J. Solids Struct.} \textbf{51}, 2254--2260. %(doi: 10.1016/j.ijsolstr.2014.03.015)

\bibitem{Colquitt2015}
Colquitt DJ, Craster RV, Makwana M. 2015. High frequency homogenisation for elastic lattices. \textit{Q. J. Mech. Appl. Math.} \textbf{68}, 203--230. %(doi: 10.1093/qjmam/hbv005)

\bibitem{Love1892}
Love AEH. 1892. \textit{A Treatise on the Mathematical Theory of Elasticity}. Cambridge: Cambridge University Press.

\bibitem{Achenbach1973}
Achenbach JD. 1973. \textit{Wave Propagation in Elastic Solids}. Amsterdam, London: North-Holland Publishing Company.

\bibitem{Graff1975}
Graff KF. 1975. \textit{Wave Motion in Elastic Solids}. New York: Dover Publications Inc.

\bibitem{Carta2018}
Carta G, Jones IS, Movchan NV, Movchan AB. 2019. Wave characterisation in a dynamic elastic lattice: lattice flux and circulation. \textit{Phys. Mesomech.} \textbf{22}(2), 152--163. %(doi: 10.1134/S102995991902005X)

\bibitem{Kelvin1894}
Thomson W. 1894. \textit{The molecular tactics of a crystal}. Oxford: Clarendon Press.

\bibitem{Spadoni2009}
Spadoni A, Ruzzene M, Gonella S, Scarpa F. 2009. Phononic properties of hexagonal chiral lattices. \textit{Wave Motion} \textbf{46}, 435--450. %(doi: 10.1016/j.wavemoti.2009.04.002)

\bibitem{SpaRuz2012}
Spadoni A, Ruzzene M. 2012. Elasto-static micropolar behavior of a chiral auxetic lattice. \textit{J. Mech. Phys. Solids} \textbf{60}, 156--171. %(doi: 10.1016/j.jmps.2011.09.012)

\bibitem{BacigalupoGambarotta2016}
Bacigalupo A, Gambarotta L. 2016. Simplified modelling of chiral lattice materials with local resonators. \textit{Int. J. Solids Struct.} \textbf{83}, 126--141. %(doi: 10.1016/j.ijsolstr.2016.01.005)

\bibitem{Tallarico2016}
Tallarico D, Movchan NV, Movchan AB, Colquitt DJ. 2016. Tilted resonators in a triangular elastic lattice: Chirality, Bloch waves and negative refraction. \textit{J. Mech. Phys. Solids} \textbf{103}, 236--256. %(doi:10.1016/j.jmps.2017.03.007)

\bibitem{PalRuzzene2017}
Pal RK, Ruzzene M. 2017. Edge waves in plates with resonators: an elastic analogue of the quantum valley Hall effect. \textit{New J. Phys.} \textbf{19}, 025001. %(doi: 10.1088/1367-2630/aa56a2)

\bibitem{Ni2015}
Ni X, He C, Sun XC, Liu XP, Lu MH, Feng L, Chen YF. 2015. Topologically protected one-way edge mode in networks of acoustic resonators with circulating air flow. \textit{New J. Phys.} \textbf{17}, 053016. %(doi: 10.1088/1367-2630/17/5/053016)

\bibitem{SusstrunkHuber2015}
S\"usstrunk R, Huber SD. 2015. Observation of phononic helical edge states in a mechanical topological insulator. \textit{Science} \textbf{349}, 6243, 47--50. %(doi: 10.1126/science.aab0239)

\bibitem{Brun2012}
Brun M, Jones IS, Movchan AB. 2012. Vortex-type elastic structured media and dynamic shielding. \textit{Proc. R. Soc. A} \textbf{468}, 3027--3046. %(doi: 10.1098/rspa.2012.0165)

\bibitem{Carta2014}
Carta G, Brun M, Movchan AB, Movchan NV, Jones IS. 2014. Dispersion properties of vortex-type monatomic lattices. \textit{Int. J. Solids Struct.} \textbf{51}, 2213--2225. %(doi: 10.1016/j.ijsolstr.2014.02.026)

\bibitem{Carta2017}
Carta G, Jones IS, Movchan NV, Movchan AB, Nieves MJ. 2017. ``Deflecting elastic prism'' and unidirectional localisation for waves in chiral elastic systems. \textit{Sci. Rep.} \textbf{7}, 26. %(doi: 10.1038/s41598-017-00054-6)

\bibitem{Wang2015}
Wang P, Lu L, Bertoldi K. 2015. Topological phononic crystals with one-way elastic edge waves. \textit{Phys. Rev. Lett.} \textbf{115}, 104302. %(doi: 10.1103/PhysRevLett.115.104302)

\bibitem{Garau2018}
Garau M, Carta G, Nieves MJ, Jones IS, Movchan NV, Movchan AB. 2018. Interfacial waveforms in chiral lattices with gyroscopic spinners. \textit{Proc. R. Soc. London A} \textbf{474}, 20180132. %(doi: 10.1098/rspa.2018.0132)

\bibitem{Nash2015}
Nash LM, Kleckner D, Read A, Vitelli V, Turner AM, Irvine WTM. 2015. Topological mechanics of gyroscopic metamaterials. \textit{Proc. Natl. Acad. Sci.} \textbf{112}, 14495--14500. %(doi: 10.1073/pnas.1507413112)

\bibitem{Hughes1986}
Hughes PC. 1986. \textit{Spacecraft Attitude Dynamics}. New York: Wiley.

\bibitem{D'EleuterioHughes1987}
D'Eleuterio GMT, Hughes PC. 1987. Dynamics of gyroelastic spacecraft. \textit{J. Guidance} \textbf{10}, 401--405. %(doi: 10.2514/3.20231)

\bibitem{Yamanaka1996}
Yamanaka K, Heppler GR, Huseyin K. 1996. Stability of gyroelastic beams. \textit{AIAA J.} \textbf{34}, 1270--1278. %(doi: 10.2514/3.13223)

\bibitem{HassanpourHeppler2016a}
Hassanpour S, Heppler GR. 2016. Theory of micropolar gyroelastic continua. \textit{Acta Mech.} \textbf{227}, 1469--1491. %(doi: 10.1007/s00707-016-1573-x)

\bibitem{HassanpourHeppler2016b}
Hassanpour S, Heppler GR. 2016. Dynamics of 3D Timoshenko gyroelastic beams with large attitude changes for the gyros. \textit{Acta Astron.} \textbf{118}, 33--48. %(doi: 10.1016/j.actaastro.2015.09.012)

\bibitem{D'EleuterioHughes1984}
D'Eleuterio GMT, Hughes PC. 1984. Dynamics of gyroelastic continua. \textit{J. Appl. Mech.} \textbf{51}, 415--422. %(doi: 10.1115/1.3167634)

\bibitem{HughesD'Eleuterio1986}
Hughes PC, D'Eleuterio GMT. 1986. Modal parameter analysis of gyroelastic continua. \textit{J. Appl. Mech.} \textbf{53}, 918--924. %(doi: 10.1115/1.3171881)

\bibitem{Carta2018b}
Carta G, Nieves MJ, Jones IS, Movchan NV, Movchan AB. 2018.  Elastic chiral waveguides with gyro-hinges. \textit{Quart. J. Mech. Appl. Math.} \textbf{71}, 157--185. %(doi: 10.1093/qjmam/hby001)

\bibitem{Nieves2018}
Nieves MJ, Carta G, Jones IS, Movchan AB, Movchan NV. 2018. Vibrations and elastic waves in chiral multi-structures. \textit{J. Mech. Phys. Solids} \textbf{121}, 387--408. %(doi: 10.1016/j.jmps.2018.07.020)

\bibitem{Carta2017b}
Carta G, Jones IS, Movchan NV, Movchan AB, Nieves MJ. 2017. Gyro-elastic beams for the vibration reduction of long flexural systems. \textit{Proc. Math. Phys. Eng. Sci.} \textbf{473}, 20170136. %(doi: 10.1098/rspa.2017.0136)

\end{thebibliography}
\end{document}